\documentclass[twocolumn]{aastex7}
\usepackage{showyourwork}
\usepackage{amsmath}
\usepackage{bm}
\usepackage{listings}
\usepackage{natbib}
\usepackage{xcolor}
\usepackage{mhchem}
\usepackage{hyperref}

\definecolor{codegreen}{rgb}{0,0.6,0}
\definecolor{codegray}{rgb}{0.5,0.5,0.5}
\definecolor{codepurple}{rgb}{0.58,0,0.82}
\definecolor{backcolour}{rgb}{0.95,0.95,0.92}
\definecolor{tedcommentcolor}{HTML}{e17701}

\newcommand{\um}{$\mathrm{\mu m}$}

\defcitealias{stevenson2020}{S20}

\newcommand{\update}[1]{\textcolor{blue}{\bf #1}}
\renewcommand{\update}[1]{#1}
\newcommand{\ghurl}{\url{https://github.com/tedjohnson12/pie-pca-paper/}}
\newcommand{\zenodourl}{\url{https://doi.org/10.5281/zenodo.21399182}}

\begin{document}

\title{A New Strategy for Using Spectroscopic Phase Curves to Characterize Non-Transiting Planets}

\shortauthors{Ted M. Johnson \& Avi M. Mandell}
\shorttitle{Characterizing Non-Transiting Planets with VPIE}


\author[0000-0002-1570-2203]{Ted M. Johnson}
\affiliation{Nevada Center for Astrophysics, University of Nevada, Las Vegas, 4505 South Maryland Parkway, Las Vegas, NV 89154, USA}
\affiliation{Department of Physics and Astronomy, University of Nevada, Las Vegas, 4505 South Maryland Parkway, Las Vegas, NV 89154, USA}
\email[show]{johnst82@unlv.nevada.edu}
\correspondingauthor{Ted M. Johnson}
\author[0000-0002-8119-3355]{Avi M. Mandell}
\affiliation{NASA Goddard Space Flight Center, 8800 Greenbelt Road, Greenbelt, MD 20771, USA}
\affiliation{Sellers Exoplanets Environment Collaboration, 8800 Greenbelt Road, Greenbelt, MD 20771, USA}
\email{avi.mandell@nasa.gov}

\begin{abstract}
We introduce a new time-series analysis strategy for combined-light exoplanet spectroscopic phase curves called the Variable Planetary Infrared Excess (VPIE) method.  VPIE can be used to extract information about the planetary flux contribution without the need for the planet to transit or the use of a stellar spectral model. VPIE utilizes a linear combination of a small set of individual spectra to produce an empirical model of the stellar contribution at each time step, thereby normalizing each spectrum and leaving only an imprint of the planet's flux in the residual data. We demonstrate the effectiveness of VPIE through simulated James Webb Space Telescope (JWST) observations of three known exoplanet orbiting late-type M stars: the warm giant TOI-519 b, the warm sub-Neptune GJ 876 d, and the temperate \update{rocky planet} Proxima Centauri b. Our results indicate that though VPIE \update{is less sensitive to} very \update{efficient} heat redistribution, it can successfully distinguish between various atmospheric circulation regimes (\update{low}, moderate, or high heat redistribution) and constrain planetary radii for non-unity day-night temperature ratios. While current performance for cooler targets may be limited by JWST spectroscopic capabilities at longer wavelengths, future VPIE improvements or new instrumentation could enable characterization of potentially habitable planets. VPIE offers a promising new framework for pulling back the veil on the population of non-transiting planets around nearby M-stars that are otherwise inaccessible to current techniques.
\end{abstract}

\section{Introduction}
\label{sec:intro}
Two decades ago, the \textit{Spitzer} Space Telescope detected photons directly from a transiting exoplanet for the first time, measuring a transiting exoplanet's flux based on the dimming of combined star/planet flux as the planet passes behind its host star during the secondary planetary eclipse, and paving the way for the modern study of transiting planets in thermal emission \citep{charbonneau2005,deming2005}. In the years that followed, observations of secondary eclipses, as well as emission over a full orbital phase (known as a phase curve), have become a standard method to characterize exoplanets. Observations with \textit{Spitzer} and, more recently, with the James Webb Space Telescope (JWST) have opened a new window into the atmospheric composition and structure of gas-rich exoplanets \citep[e.g.][]{madhusudhan2011,kempton2023}, and have now begun to help constrain the presence or absence of atmospheres on small rocky worlds \citep[e.g.][]{kreidberg2019,greene2023,weinermansfield2024}.

\begin{figure*}
    \centering
    \includegraphics[width=0.8\textwidth]{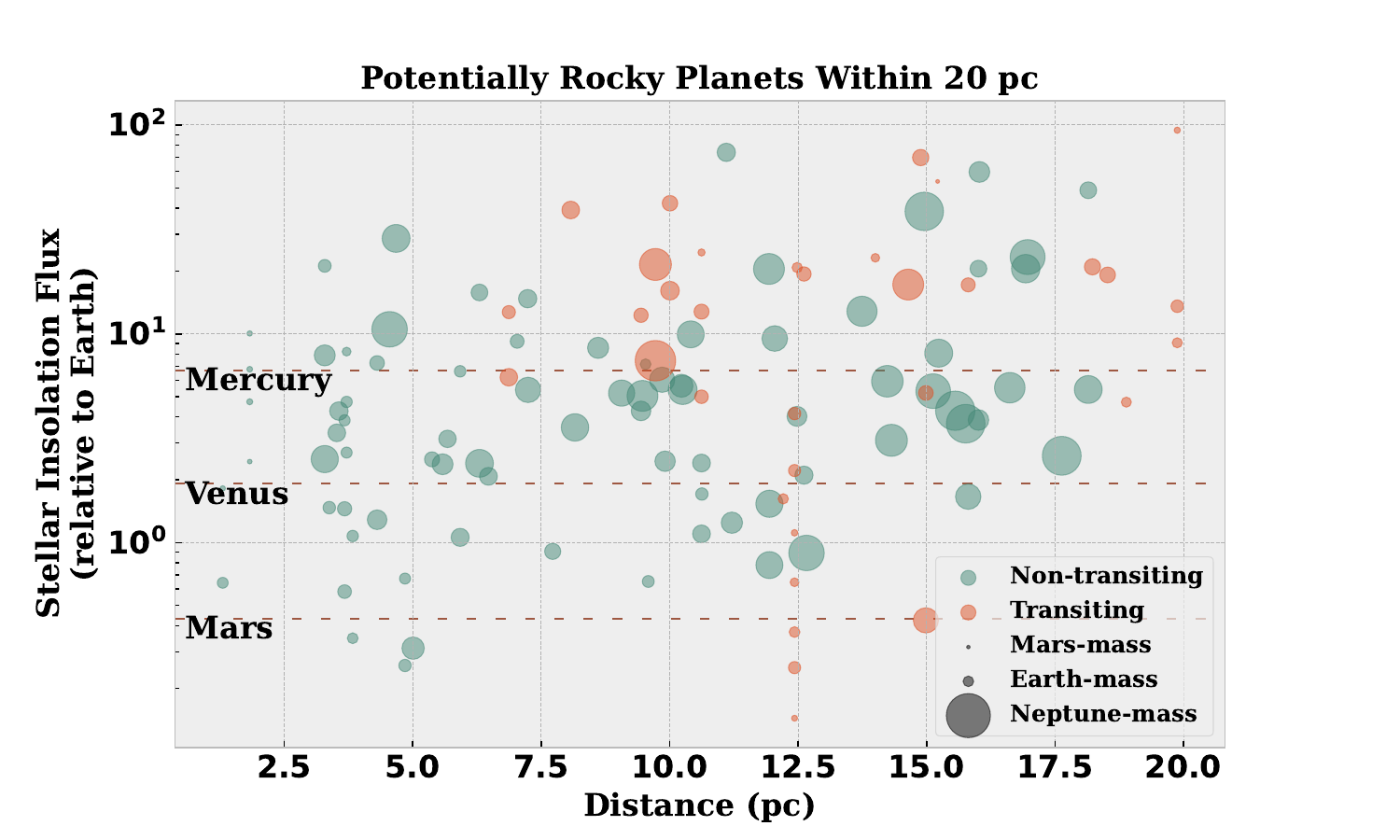}
    \caption{Known potentially-rocky exoplanets within 20 pc. The vast majority of the closest planets are non-transiting and therefore their atmospheres cannot be characterized with currently-available techniques. Data from the NASA Exoplanet Archive, accessed on %
  \input{output/nearby_mdwarfs_date.txt}\unskip\label{output/nearby_mdwarfs_date.txt}\unskip%
 \citep{nea13,christiansen2025}.
    }
    \script{nearby_mdwarfs.py}
    \label{fig:nearby_mdwarfs}
\end{figure*}

The secondary eclipse and traditional phase curve methods, however, rely on a transit geometry -- i.e., the orbital plane of the planet crossing the observer's line-of-sight to the star.  The probability of a planet producing a transit is approximately $R_*/a$ - a ratio which is $\sim 10\%$ for a Hot Jupiter around a G-dwarf, but $\sim 1\%$ for a planet in the Habitable Zone (HZ) of a late M-dwarf.  Unfortunately, the large majority of known planets around nearby M-dwarfs, primarily discovered using radial velocity (RV) measurements, are non-transiting (Figure \ref{fig:nearby_mdwarfs}); within 10 pc, there are no cool transiting planets, leaving Gliese-12 b and TRAPPIST-1 e/f (at of distance of $\sim$ 12 pc) and LHS~1140 c (at $\sim$ 15 pc) as the best candidates to characterize atmospheres in the HZ using secondary eclipse measurements. However, predictions suggest obtaining sufficient signal-to-noise would require 10 s to 100 s of eclipses for the smallest, coolest planets and may be infeasible in the lifetime of JWST \citep{morley2017}.



\update{
    This reality has led to a renewed interest in measuring planetary flux in combined (star-planet) light without the use of a secondary eclipse for baselining the stellar flux -- with a primary focus on detecting thermal phase variations.  The first detection of thermal phase variations for a non-transiting planet was achieved for the super-Jupiter $\mathrm{\upsilon}$ Andromedae b using Spitzer measurements at 24$\;\mathrm{\mu m}$ \citep{harrington2006,crossfield2010}, providing the first firm confirmation of day-night temperature variations on highly irradiated exoplanets. Early work predicted spectroscopic signals from non-transiting terrestrial planets both with \citep{selsis2011} and without \citep{maurin2012} atmospheres but with relatively simplified observational assumptions; in particular, the challenge of stellar variability was raised as a potential limiting factor for success.
}   
\update{
    The topic received new interest with the detection of a potentially habitable planet, Proxima Centauri b \citep[PCb,][]{anglada-escude2016}, orbiting the nearest star in 2016.  PCb provides an optimal planet-to-star ratio (due to the small size of the parent star) and photon-driven noise limit due to its proximity to Earth -- but does not transit its star. Shortly after its discovery, several teams (\citet{turbet2016, kreidberg2016}) proposed that phase variations in thermal emission could be detected with the then-future observatory JWST. Those authors drew two important conclusions: 1) that planetary emission could dominate over photon noise in a (relatively short) 24 hour exposure, and 2) thermal emission from the day and night sides could together constrain planetary radius, Bond albedo, and atmospheric heat redistribution. The detection of an atmosphere (inferred from advection of heat from day to night) on PCb would be revolutionary; to date, there are no observational constraints as to whether HZ M-Earths can even host atmospheres -- a major question due to the intense irradiation from high-energy photons for planets around highly active M-stars \citep[e.g.][]{airapetian2017,pass2025}.
}

\update{
    However, as noted previously, overcoming photon noise is unlikely to be the most challenging hurdle in detecting an M-Earth atmosphere; \citet{kreidberg2016} note that stellar variability may be on the order of 1\%, orders of magnitude higher than the planetary signal. A decade later, it is clear that biases introduced by stellar heterogeneity (i.e. variability or the transit light-source effect, \citealp{rackham2018}) are consistently a major challenge in characterizing exoplanet atmospheres \citep{moran2023,sotzen2025}.
}

\update{
    More recently, there has been progress towards the eventual characterization of non-transiting terrestrial-planet atmospheres.
} \citet{stevenson2020} introduced a new method to measure the thermal emission of an exoplanet that does not require a secondary eclipse. Called Planetary Infrared Excess (PIE), this technique is akin to the way that the radially-dependent emission from circumstellar disks can be characterized by measuring excess infrared light emitted by a stellar system \update{ (e.g. \citet{meyer2008}).}
\update{
    This strategy was explored explicitly in relation to the detection of planets around white dwarfs with JWST \citep{limbach2022}, and led to the discovery of planetary candidates from thermal excess emission with JWST/MIRI \citep{limbach2024,limbach2025}. \citet{mayorga2023} found that PIE observations could provide measurements complementary to transits and RVs to better constrain the properties of rocky multi-planet systems, and \citet{lustig-yaeger2025} then empirically validated the PIE technique using JWST/NIRSpec observations of both night-side and day-side spectra of the transiting hot Jupiter WASP-17 b, showing equivalence between the PIE-derived results and traditional eclipse lightcurve fitting. In parallel, \citet{mandell2022} built upon the technique to propose a PIE-optimized space mission capable of characterizing 20--40 nearby rocky planets in thermal emission.
}

\update{The slow-but-steady progress towards non-transiting terrestrial planets has been hampered by a major challenge for PIE: the infrared excess from planetary thermal emission} is very small -- on the order of $10^{-5} - 10^{-4}$ with respect to the star, and therefore extreme care is needed to successfully extract the planetary flux from the underlying stellar contribution. Unfortunately, the key impediment to reaching this level of precision with the original PIE method is that it requires a physics-based stellar model to be accurate to high precision across a broad wavelength range. Recent JWST studies have found that even the best models can be reliable only to within $\sim 1\%$ \citep{moran2023}. Additionally, it has been found that inaccuracies in stellar models lead to significant systematic errors in the absolute flux estimates of eclipsing planets and the worst inaccuracies are in models of M-dwarfs, which host most of the nearby planets known today \citep{fauchez2025}.

\update{Accurate characterization of rocky M-dwarf planets will either require more advanced stellar models, or techniques that do not rely on the use of physics-based models. For example, \citep{deming2026} suggests that the broadband color evolution of the combined star-planet flux could be used to infer properties of the planets by intrinsically accounting for stellar flux amplitude variations}.
In this study, we introduce an adaptation to the original PIE method called the Variable Planetary Infrared Excess (VPIE) method that extends the concepts of PIE in the temporal domain to separate planetary and stellar flux variability without the need for a physics-based stellar model. We describe the basic methodology of the VPIE normalization scheme in Section \ref{sec:vpie}, and then apply that methodology to three simulated JWST observations in Section \ref{sec:examples}. In Section \ref{sec:disc} we discuss potential limitations of this method and the prospect for future observatories which would be better suited for VPIE observations. Finally, we summarize our findings in Section \ref{sec:conclusion}
\section{The VPIE Method}
\label{sec:vpie}

VPIE is a PIE time-series analysis method that aims to measure the thermal emission of a planet as it orbits its host. This is done by treating each spectrum of the target as an instantaneous measurement of multiple emitting components of the star and the planet, and the set of observations as a combination of many different measurements of these components over time. This method is most sensitive to variations in the planetary spectrum as it rotates from day-side to night-side facing an observer. 

VPIE relies on a scenario where a planetary system is observed in combined light over a broad wavelength range, such that the spectrum can be divided into two portions:
\begin{enumerate}
    \item The short wavelength (SW) region; at these wavelengths, the photon noise from the star dominates over the planetary spectrum, and the planet effectively contributes zero flux.
    \item The long wavelength (LW) region; here, the flux from the planet rises above the photon noise, and so, in theory, it has a detectable contribution to the total flux.
\end{enumerate}

The goal of VPIE is to utilize a linear combination of a small set of individual spectra in the time series to produce a ``model'' of the stellar contribution at each time step, thereby normalizing each spectrum and leaving only an imprint of the planet's flux in the residual data. A planetary phase curve model can then be fit to these residuals to infer information about the planet, such as its day-night temperature difference, emitting area (i.e. size), spectral absorbers, etc. In the subsequent subsections, we outline the rationale and mathematical formulation for our normalization scheme.

\subsection{Spectral Normalization}\label{sec:change-basis}

This section describes the core spectral-normalization scheme employed by VPIE. We will describe the mathematics mostly using vectors (written in {\bf \em bold italics}) to represent spectra at a single epoch, but in reality much of the actual math is much more concise when written in matrix form. For a detailed description of the mathematics involved in implementing VPIE, see Appendix \ref{ap:math-desc}. As stated above, to construct the normalization model, we will only work with the SW spectral region, which contains only stellar flux and no information about the planet.

There are many sources of stellar variability that ultimately arise from both the overall rotation of the star as well as the evolution of various physical features on the stellar surface. For example, stellar rotation can change the fraction of the surface visible to the observer that is covered by star spots -- and star spots could in theory have a range of effective temperatures. Similarly, convection-induced granulation produces stochastic variations of a low-temperature inter-granule component \update{\citep[e.g.,][]{kallinger2014,coulombe2025}}. We call such sources of variability {\em component mixing variability}. Importantly, a star whose variability is solely due to component mixing will have a spectrum that is well described by a linear combination of other spectra dominated by each component. Each observed spectrum is an element of the vector space spanned by the spectral contributions of each stellar component.

It is not the goal of VPIE to learn about the physical processes governing stellar variability, so the characteristics of the individual stellar spectra themselves are not important for this application. Instead, we choose individual epochs from the observation to act as {\em basis spectra}. The basis is chosen using the methods described in Appendix \ref{ap:choose-basis}, which ultimately test multiple combinations of basis spectra to minimize the Bayes Information Criterion \citep[BIC,][]{schwarz1978} or Akaike Information Criterion \citep[AIC,][]{akaike1974}, both of which are metrics that balance model error with complexity.

\begin{figure}
    \centering
    \includegraphics[width=0.35\textwidth]{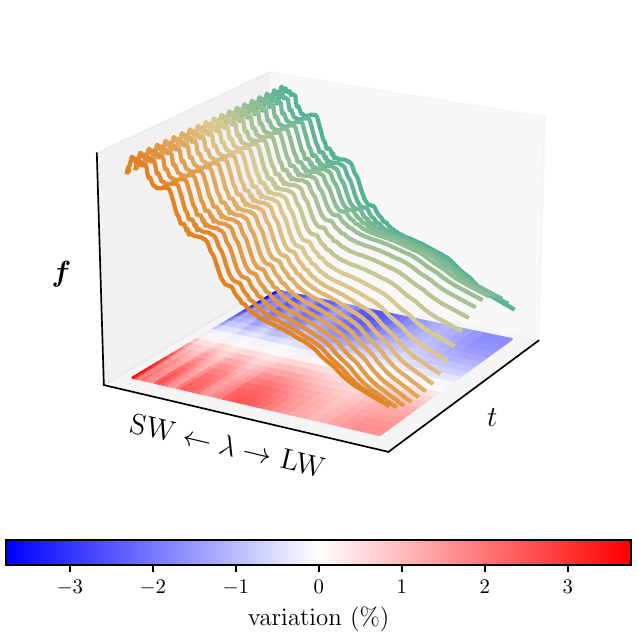}
    \includegraphics[width=0.32\textwidth]{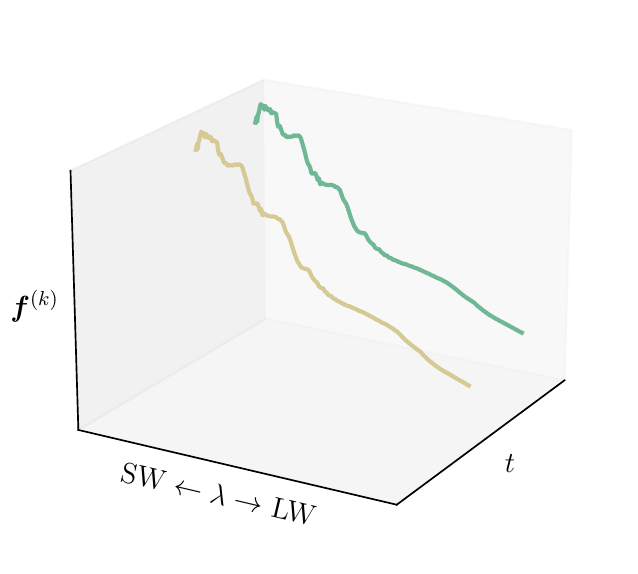}
    \includegraphics[width=0.35\textwidth]{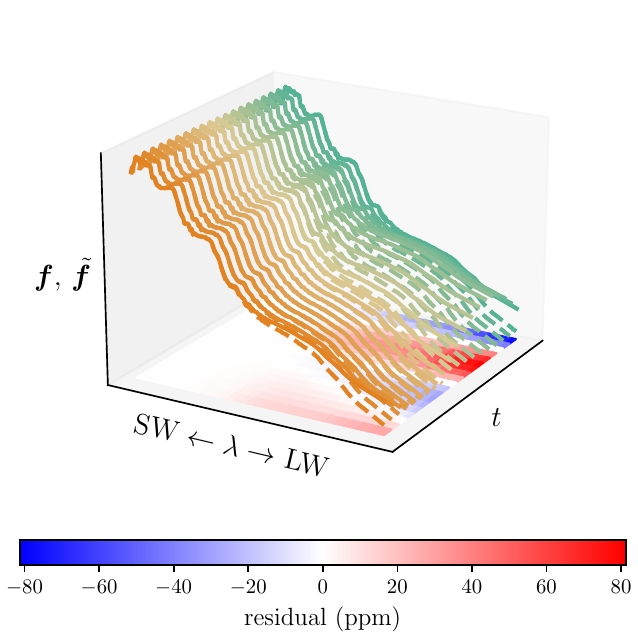}
    \caption{Basis spectra reconstruction. {\bf Top:} Spectral time-series comprising a VPIE observation. The planetary spectrum has been multiplied by $10^4$ for visibility; the sinusoidal shape of the variability caused by the planet can be clearly seen in the LW portion of the spectrum. The SW stellar spectrum also varies with time; the red-blue image on the $x$-$y$ plane shows variations from the mean spectrum. {\bf Middle:} The best basis for the dataset shown in top panel, found by minimizing the BIC. {\bf Bottom:} Reconstruction of the dataset shown in top panel using only the basis spectra (dashed lines), plotted along with the original data (solid lines). Note that solid and dashed lines converge in the SW portion of the spectrum, but in the LW region the spectra are significantly different; this is because the fitting procedure was not given any information about the time-series behavior of the planet. The planet's imprint on the residuals is shown in red and blue on the $x$-$y$ plane; note the difference in the \update{color bar scales}.}
    \label{fig:3d}
    
    \script{fig3d_final.py}
\end{figure}

Once a set of basis spectra have been chosen (which can be written $\{\bm{f}^{(1)},\,\bm{f}^{(2)},\,\dots,\,\bm{f}^{(q)}\}$), we {\em reconstruct} the set of observed spectra using a linear combination of the bases. This is done via weighted least-squares fitting of the SW region of each spectrum. Each spectrum in the time-series can now be represented by a few coefficients which describe how to construct it from the basis spectra. Recall that there is no planetary information in the SW spectrum, so by choosing the basis and performing this fitting using only SW data, we have a model for the stellar variability which is agnostic to the planet. Ultimately, as shown in Figure \ref{fig:3d}, we reconstruct the whole dataset (SW and LW) using the coefficients from the least-squares procedure. The result is \update{an empirical} model observation ($\tilde{\bm{f}}$) that fits very well at short (star-only) wavelengths, but significantly differs from the original data at long wavelengths where the planet has appreciable flux.

\subsection{Computing Residuals}
\label{sec:residuals}


After using least-squares to fit the basis spectra to our observation, we subtract the reconstruction from the original data to compute residuals ($\bm{f} - \tilde{\bm{f}}$), as shown in the bottom panel in Figure \ref{fig:3d}. Because we did not include any LW data in our least-squares fit, the reconstructed model is unaware of the planetary variations that occur at these wavelengths. The residuals are equal to the sum of an error vector ($\bm{\epsilon}$) and a planetary variability vector ($\bm{\delta}$), written
\begin{equation}\label{eq:def-residuals}
    \bm{f}_i - \tilde{\bm{f}}_i = \bm{\epsilon}_i + \bm{\delta}_i\, ,
\end{equation}
for time index $i$. Therefore, the LW residuals encode any planetary signal that is not correlated with stellar variability at short wavelengths, with the strength of the residual signal inversely proportional to the magnitude of correlation. In this sense, the planetary signal extraction actually benefits from a larger amplitude of stellar variability, as long as it is uncorrelated with the time-dependent planetary signal.

Specifically, if we have written our reconstructed spectrum as
\begin{equation}\label{eq:def-frec}
    \tilde{\bm{f}} = \sum_{k=0}^q a_{k}\,\bm{f}^{(k)}\, ,
\end{equation}
then Appendix \ref{ap:residuals} shows that the residual due to the planet is
\begin{equation}\label{eq:def-delta}
    \bm{\delta} = \bm{f}_\text{p} - \sum_{k=0}^q a_{k}\,\bm{f}_{\text{p}}^{(k)}\, ,
\end{equation}
where $\bm{f}_\text{p}$ is the intrinsic planetary spectrum and $a_k$ is the coefficient for the $k$th basis spectrum computed via least-squares. The raised index $(k)$ denotes that we are summing the $k$th basis spectrum, not the $k$th spectrum in the time-series. 

Most importantly, the quantity $\bm{\epsilon} + \bm{\delta}$ is measurable, and Section \ref{sec:compute-delta} shows that $\bm{\delta}$ can be computed from a planetary model. The error term $\bm{\epsilon}$ arises because the basis spectra are not expected to be a perfect fit (due to scatter or non-component mixing variability). However, if it is mostly due to photon noise, then the residuals should have a mean value of $\bm{\delta}$ and can be used to make inferences about the properties of the planet.

A more complete notation would write equations (\ref{eq:def-frec} \& \ref{eq:def-delta}) with the time index $i$ included (i.e. $\bm{\delta}\rightarrow\bm{\delta}_i$, $a_k\rightarrow a_{ik}$, etc.), but this is not necessary for this simplified description. See Appendix \ref{ap:residuals} for more detail.

\subsection{Planetary Model Inversion}\label{sec:compute-delta}

The vector $\bm{\delta}$ is the \update{component of the residual which encodes} the intrinsic variability of the planet. Suppose we have a specific model that computes a planetary spectroscopic phase curve ($\bm{f}_{\text{p},\,1},\, \bm{f}_{\text{p},\,2},\,\dots$) and we would like to compare it to the residuals from an observation. The definition of $\bm{\delta}$ in equation (\ref{eq:def-delta}) tells us exactly how to compute it from the model, but we discuss here the meaning of the indices and coefficients in this equation. Because this is not the true value of $\bm{\delta}$, we will write it as a function of model parameters: $\bm{\delta}(\theta)$.

Recall that minimizing BIC (or AIC) in Section \ref{sec:change-basis} resulted in a set of basis spectra, described by their indices in the spectral time-series. Those same indices describe the planetary spectral basis, denoted by the index $k$ in equation (\ref{eq:def-delta}). If $\bm{f}_\mathrm{p}(\theta)$ is the planet flux predicted by a model, then that same model can predict the planet's imprint in the $\bm{f}-\tilde{\bm{f}}$ residuals:
\begin{align} \label{eq:model-residual}
    \bm{\delta}(\theta) = \bm{f}_\mathrm{p}(\theta) - \sum_{k=0}^q a_k \bm{f}_\mathrm{p}^{(k)}(\theta)\, .
\end{align}

We can compare the (observed) $\bm{\delta}$ to various $\bm{\delta}(\theta)$ to infer the model parameters that best describe the planet. \update{Figure 3 uses a 1-dimensional example to visually demonstrate how various models can be compared to a VPIE measurement in order to infer planetary properties. The LW residuals ($\bm{\delta}$, eq. \ref{eq:def-delta}) are shown as the blue line with shading representing uncertainty; they deviate from zero due to the planetary thermal emission modulated by the stellar variability signal. The fact that the residuals are mostly negative is a result of the specific basis spectra chosen and the nature of both stellar and planetary variability with respect to those basis spectra; in general, the planet's imprint on the residuals produces both positive and negative values of similar magnitude. 
Various phase curve models (with parameters $\theta$) can be compared to this data by computing the expected residual pattern ($\bm{\delta}(\theta)$, eq. \ref{eq:model-residual}). The properties of the planet can then be inferred based on each model's goodness-of-fit}
\update{
\begin{align}
    \chi^2_\mathrm{red} =\frac{1}{K} \sum\frac{(\bm{\delta} - \bm{\delta}(\theta))^2}{\bm{\sigma}^2},
\end{align}
}
\update{where $\delta$ is defined by eq. (\ref{eq:def-delta}), $\delta(\theta)$ by eq. (\ref{eq:model-residual}), and $K$ is the number of degrees of freedom, in this case the number of wavelength channels ($n$) times the number of temporal bins ($m$) minus the two parameters we fit for: $n\times m -2$. Note that the basis-vector reconstruction procedure introduces additional noise, which is mostly mitigated by binning in time and wavelength after the reconstruction; additionally, we inflate the error bars by a factor so that the minimum $\chi^2_\mathrm{red}=1$; scaling values ranged from 1.05--2.3 based on the dataset.}

\begin{figure}
    \includegraphics[width=0.5\textwidth]{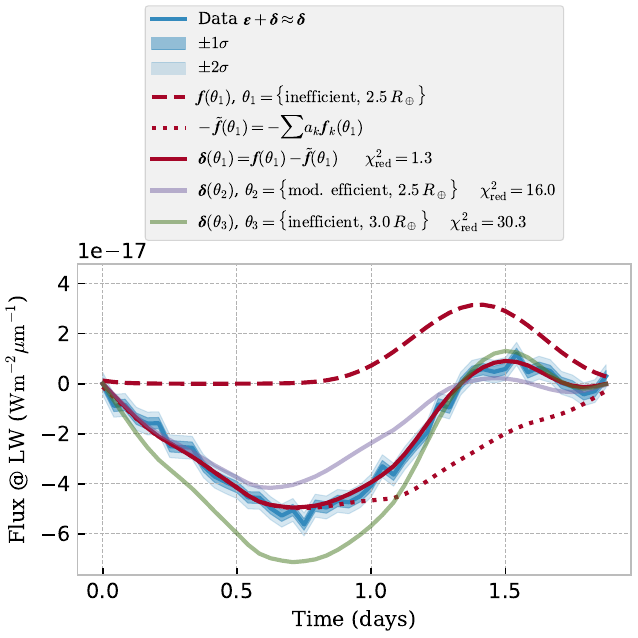}
    \caption{
        \update{
        Example of planetary properties inference with 1-dimensional data. The properties of the planet in the simulated observation (blue with shading) can be inferred by comparing to various models (solid lines). The red dashed line is a phase curve model with input parameters $\theta_1$. As an intermediate step we show the phase curve reconstruction ($-\bm{\tilde{f}}$) which traces stellar activity; this is the dotted line, note that we multiplied by a factor of $-1$ so that it may be added to the dashed line by eye. The sum of the dashed and dotted lines is the planet model's residual signal ($\bm{\delta}(\theta_1)$, solid red line). Residual signals for two additional models $(\theta_2,\,\theta_3)$ are also shown. Each residual is compared to the data using the $\chi^2_\mathrm{red}$ metric, and as a result we may infer that the data supports a planet with properties similar to $\theta_1$.
        }
    }
    \label{fig:chi2demo}
    \script{chi2_test.py}
\end{figure}

\update{Our analysis purposefully does not utilize a sophisticated statistical inference tool such as Bayesian inference techniques or optimal estimation. Our goal here is simply to examine how the ``signal'' of the planetary phase curve changes when varying two of the dominant factors driving the amplitude of the phase-dependent variations, planetary radius and day-night temperature ratio, over reasonable values; we then compare the signal to the uncertainty expected under specific realistic observing scenarios.}

\section{Simulated VPIE Analyses for Known Exoplanets}\label{sec:examples}

In this section we explore three proof-of-concept demonstrations of the VPIE method using simulated observations, walking through the steps outlined in previous sections, in order to quantify the fundamental constraints on planetary radius and atmospheric circulation that can be retrieved for a specific known exoplanet target under a specific observing scenario.

\update{In the following sub-sections we examine three potential targets for characterization with VPIE: the warm giant TOI-519 b, the warm sub-Neptune GJ 876 d, and the potentially-habitable terrestrial planet Proxima Centauri b.}
For each target, we model the stellar and planetary time series based on observed constraints or reasonable physical assumptions; similarly, we assume the observatory characteristics of either existing JWST instrumentation or a reasonable extension of those characteristics. Furthermore, we model three planetary atmosphere scenarios: 1) No atmosphere/highly-inefficient heat redistribution, where the night-side of the planet is very cold compared to the day-side \citep{nguyen2020,hammondm2025}, 2) moderately-efficient heat transport, which is expected for highly-irradiated giant planets \citep{perna2012,perez-becker2013,komacek2016} as well as rocky planets with thin to moderately-thick atmospheres, and 3) highly-efficient heat transport, where the day- and night-sides are the same or nearly the same temperature. This is expected on cooler giant planets and terrestrial planets with thick atmospheres \citep{hammond2017}.

For each case, we first generate a simulated combined-light dataset with both stellar and planetary flux using the Variable Star PhasE Curve code \citep[\texttt{VSPEC, }][]{johnson2025}, which uses the Planetary Spectrum Generator's Global Emission Spectra application \citep[PSG/GlobES][]{villanueva2018,kofman2024} to create combined-light spectral timeseries including the photon-noise contribution under specified observatory characteristics. \update{\texttt{VSPEC} generates simplified 3D atmosphere models based on instellation and a dimensionless ``thermal inertia'' parameter, \citep[$\epsilon$, see][]{cowan2011}. Herein we translate their $\epsilon$ into a more-intuitive night/day temperature ratio ($T_\mathrm{night}/T_\mathrm{day}$), where $T_\mathrm{night}$ and $T_\mathrm{day}$ are the minimum and maximum equatorial temperatures, respectively. For computational tractability, these simulations have been done at low spatial resolution of the planetary surface; using higher-resolution simulated data to better resolve hot spots on the dayside would increase the peak planetary flux but not significantly alter our results.}

\update{
We then analyze the combined-light phase curve, using VPIE to isolate the planet's variability ($\bm{\epsilon}+\bm{\delta}$, see eq. \ref{eq:def-residuals}).}
Finally, we generate a full grid of planetary forward models with \update{various radii}  and heat-redistribution \update{efficiencies, compute $\bm{\delta}(\theta)$ for each model, and evaluate its goodness-of fit to the data. 
}

\subsection{Example 1: A Transiting Warm Giant Planet in the NIR}\label{subsec:toi519}

Giant exoplanets around M-stars (GEMS) are the optimal candidates for demonstrating VPIE observations because the combination of a large planet and small star maximizes the planet/star flux ratio. The discovery of GEMS has been catalyzed by the combination of TESS and ground-based RV studies of nearby M-dwarfs, and there are now more than 15 planets more massive than Saturn discovered around M-stars with $T_\mathrm{eff}<3500\;\mathrm{K}$ (\update{NASA} Exoplanet Archive; \citealt{nea13,christiansen2025}). In particular, GEMS with equilibrium temperatures below $\sim800\;\mathrm{K}$ would produce \update{negligible} thermal flux below $1.5\;\mathrm{\mu m}$, and would therefore be amenable to VPIE analysis using observations spanning the $0.6 - 5\;\mathrm{\mu m}$ NIR wavelength range covered by several JWST instruments (NIRSpec, NIRCam, NIRISS).

To determine the effectiveness of VPIE for a GEMS target with JWST under realistic observing conditions, we simulate a JWST phase-curve observation of the warm ($T_\mathrm{eq}\approx 750\; \mathrm{K}$) Jupiter-sized exoplanet TOI-519 b, which transits its mid-M host star with $T_\mathrm{eff}\approx 3300\; \mathrm{K}$ \citep{parviainen2021,hartman2023,kagetani2023}. TOI-519 b is the largest planet detected so far that meets the stellar and planetary temperature criteria above, and is therefore the optimal test case.

Table \ref{tab:parameters} gives the VSPEC model parameters used in this simulation.  The bulk stellar and planetary parameters were taken from the most recent values listed in the NExSci Exoplanet Archive; individual references are listed next to each value. We defined the stellar photospheric temperature to be $3413\;\mathrm{K}$, with a spot temperature of $2700\;\mathrm{K}$ and a spot coverage fraction of 11\% so that the total flux matches the reported $T_\mathrm{eff}$ of $3354\;\mathrm{K}$ \citep{kagetani2023}; these stellar spot parameters are selected based on those inferred for similar M3.5V GJ 486 via the transit light-source effect \citep{moran2023}. The planetary $T_\mathrm{eq}$ of \update{$767\;\mathrm{K}$} is based on an assumption of a geometric albedo of zero, which is clearly a lower limit but similar to values measured for hot Jupiters \citep{krenn2023}.  Observation parameters are based on values for the JWST NIRSpec/PRISM instrument, with VPIE-specific parameters such as the time bin size and cut-off wavelength optimized for this target.

%
  \begin{table*}
\centering
\begin{tabular}{ccccc}
\hline
Quantity & TOI-519 b & GJ 876 d & PCb/JWST & PCb/Future Observatory \\
\multicolumn{5}{c}{~\leaders\vrule height 2.5pt depth -2pt \hfill \null~Stellar Parameters~\leaders\vrule height 2.5pt depth -2pt \hfill \null~} \\
Stellar Effective Temperature & $3354 \; \mathrm{K}$$^a$ & 3293 K$^b$ & 2900 K$^c$ & - \\
Stellar Radius & $0.36 \; \mathrm{R_{\odot}}$$^a$ & $0.37 \; \mathrm{R_{\odot}}$$^b$ & $0.141 \; \mathrm{R_{\odot}}$$^c$ & - \\
Stellar Rotation Period & $4 \; \mathrm{d}$$^\dagger$ & $95 \; \mathrm{d}$$^d$ & $90 \; \mathrm{d}$$^c$ & - \\
Photosphere Temperature & $3407 \; \mathrm{K}$ & $3342 \; \mathrm{K}$ & $2962 \; \mathrm{K}$ & - \\
Spot Temperature & $2700 \; \mathrm{K}$$^\dagger$ & $2700 \; \mathrm{K}$$^\dagger$ & $2600 \; \mathrm{K}$$^\dagger$ & - \\
Spot Coverage Fraction & 0.1$^\dagger$ & 0.2$^\dagger$ & 0.2$^\dagger$ & - \\
\multicolumn{5}{c}{~\leaders\vrule height 2.5pt depth -2pt \hfill \null~Planet Parameters~\leaders\vrule height 2.5pt depth -2pt \hfill \null~} \\
Planet Radius & $1.03 \; \mathrm{R_{\rm J}}$$^a$ & $2.5 \; \mathrm{R_{\oplus}}$$^\dagger$ & $1 \; \mathrm{R_{\oplus}}$$^\dagger$ & - \\
Planet Mass & $147 \; \mathrm{M_{\oplus}}$$^a$ & $7.49 \; \mathrm{M_{\oplus}}$$^d$ & $1 \; \mathrm{M_{\oplus}}$$^\dagger$ & - \\
Planet $T_\mathrm{eq}$ & $767 \; \mathrm{K}$ & $664 \; \mathrm{K}$ & $243 \; \mathrm{K}$ & - \\
Semimajor Axis & $0.0159 \; \mathrm{AU}$$^a$ & $0.0218 \; \mathrm{AU}$$^d$ & $0.04856 \; \mathrm{AU}$$^c$ & - \\
Orbital Period & $1.265 \; \mathrm{d}$$^a$ & $1.938 \; \mathrm{d}$$^d$ & $11.187 \; \mathrm{d}$$^c$ & - \\
Eccentricity & 0.0$^\dagger$ & 0.1$^d$ & 0.0$^c$ & - \\
Initial Phase & $90\mathrm{{}^{\circ}}$$^\dagger$ & $90\mathrm{{}^{\circ}}$ & $90\mathrm{{}^{\circ}}$$^\dagger$ & - \\
Distance & $115 \; \mathrm{pc}$$^e$ & $4.672 \; \mathrm{pc}$$^e$ & $1.302 \; \mathrm{pc}$$^e$ & - \\
Inclination & $88.9\mathrm{{}^{\circ}}$$^a$ & $53.19\mathrm{{}^{\circ}}$$^d$ & $80\mathrm{{}^{\circ}}$$^\dagger$ & - \\
Mean Molecular Weight & 2$\;\mathrm{g\;cm^{-3}}$$^\dagger$ & 28$\;\mathrm{g\;cm^{-3}}$$^\dagger$ & 28$\;\mathrm{g\;cm^{-3}}$$^\dagger$ & - \\
Albedo & 0.0$^\dagger$ & 0.0$^\dagger$ & 0.0$^\dagger$ & - \\
\multicolumn{5}{c}{~\leaders\vrule height 2.5pt depth -2pt \hfill \null~Instrument Parameters~\leaders\vrule height 2.5pt depth -2pt \hfill \null~} \\
Aperature & $5.6 \; \mathrm{m}$ & $5.6 \; \mathrm{m}$ & $5.6 \; \mathrm{m}$ & $0.5 \; \mathrm{m}$, $2 \; \mathrm{m}$, - \\
Observation Length & $1.3 \; \mathrm{d}$ & $1.9 \; \mathrm{d}$ & $11.2 \; \mathrm{d}$ & $179 \; \mathrm{d}$, -, - \\
Observation Cadence & $30 \; \mathrm{min}$ & $60 \; \mathrm{min}$ & $4 \; \mathrm{h}$ & - \\
Short Wavelength & $0.6 \; \mathrm{\mu m}$ & $0.6 \; \mathrm{\mu m}$ & $5 \; \mathrm{\mu m}$ & $1 \; \mathrm{\mu m}$ \\
Long Wavelength & $5 \; \mathrm{\mu m}$ & $5 \; \mathrm{\mu m}$ & $12 \; \mathrm{\mu m}$ & $18 \; \mathrm{\mu m}$ \\
Resolving Power & 100 & 100 & 100 & 50 \\
SW & $<\;$$2.5 \; \mathrm{\mu m}$ & $<\;$$1.5 \; \mathrm{\mu m}$ & $<\;$$7 \; \mathrm{\mu m}$ & $<\;$$3 \; \mathrm{\mu m}$ \\
LW & $>\;$$4 \; \mathrm{\mu m}$ & $>\;$$3.5 \; \mathrm{\mu m}$ & $>\;$$10 \; \mathrm{\mu m}$ & $>\;$$8 \; \mathrm{\mu m}$ \\
\multicolumn{5}{c}{~\leaders\vrule height 2.5pt depth -2pt \hfill \null~Inference Grid~\leaders\vrule height 2.5pt depth -2pt \hfill \null~} \\
Radius & $0.1-2.1\,R_\mathrm{J}$ & $0.1-8.0\,R_\oplus$ & $0.1-3.2\,R_\oplus$ & - \\
$T_\mathrm{night}/T_\mathrm{day}$ & $0.05-0.99$ & $0.05-0.99$ & $0.05-0.99$ & - \\
\hline
\end{tabular}
\caption{Simulation Parameters. Values only listed for PCb/Future Observatory if different from PCb/JWST. $^\dagger$assumed; $^a$\citet{kagetani2023}; $^b$\citet{rosenthal2021}; $^c$\citet{faria2022}; $^d$\citet{nelson2016}; $^e$\citet{gaiacollaboration2020}} 
\label{tab:parameters}
\end{table*}\unskip\label{output/tab.txt}\unskip%

\begin{figure*}
    \gridline{
        \fig{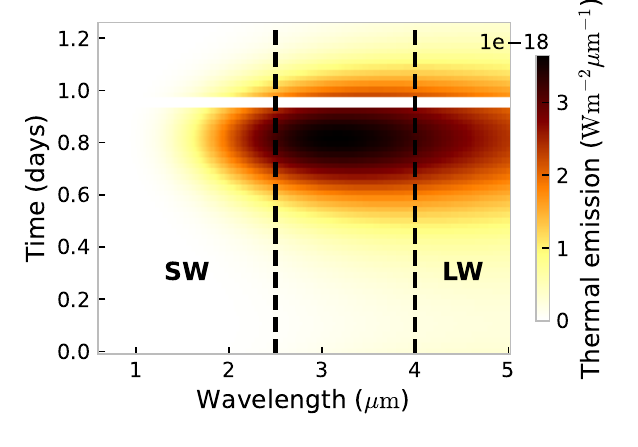}{0.5\textwidth}{\bf (a)}
        \fig{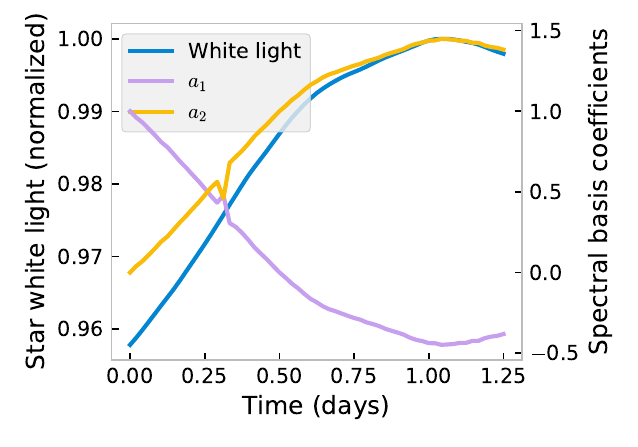}{0.5\textwidth}{\bf (b)}
    }
    \gridline{
        \fig{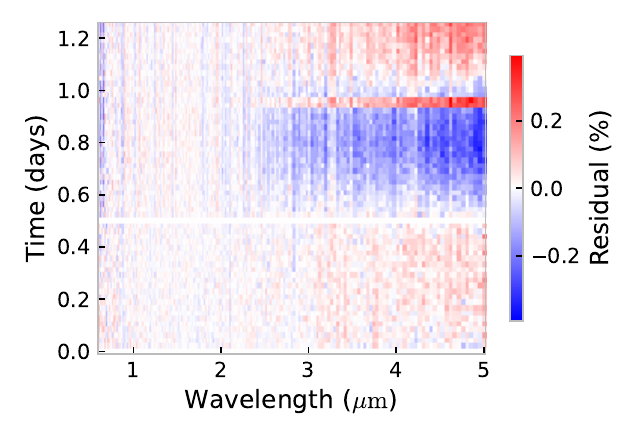}{0.5\textwidth}{\bf(c)}
        \fig{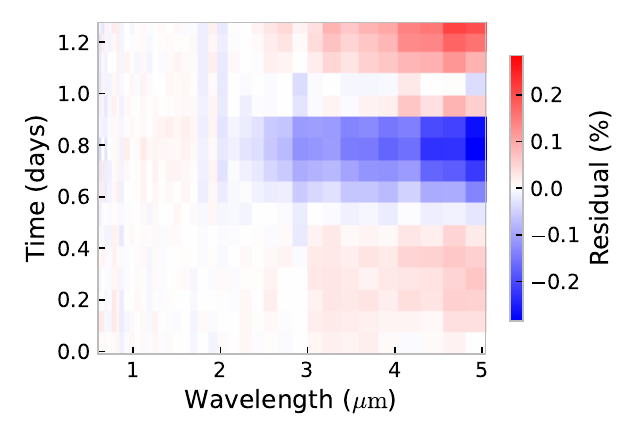}{0.5\textwidth}{\bf(d)}
    }
    \caption{
        PIE phase curve and VPIE analysis of the warm giant planet TOI-519 b. {\bf a)} Thermal emission phase curve with $T_\mathrm{night}/T_\mathrm{day} = 0.5$ over one orbit in the spectra range $0.6-5\;\mathrm{\mu m}$. {\bf b)} Fit to the stellar variability. Left axis (blue line) shows the normalized white light curve of the star over the observation. The right axis shows the best-fit coefficients of the basis spectra ($a_k$ in equation (\ref{eq:def-frec})). {\bf c)} VPIE residual pattern. This is the signal which can be matched with a model to infer planetary properties. {\bf d)} Binned VPIE residual pattern. Same as {\bf (c)}, but binned by 6 pixels in wavelength and 4 pixels in time in order to remove correlated noise.
    }
    \label{fig:toi519}
    \script{toi519_plot.py}
\end{figure*}

Figure \ref{fig:toi519} shows the data input into this example and the results of the VPIE analysis. Panel (a) shows the planetary thermal emission only, over the full orbital period and wavelength range; it demonstrates how wavelengths shorter than $\sim2\;\mathrm{\mu m}$ are essentially free of planetary flux. Panel (b) shows the evolution of the coefficients for the two basis-vector spectra used to create the normalization model (designated as $a_1$ and $a_2$), as well as the normalized values of the summed ``white-light'' time-series; the BIC analysis determined that a linear combination of two basis vectors produced the best fit to the SW data. Finally, panels (c) and (d) demonstrate the final, measurable product of the VPIE technique: the residuals of $\bm{f}-\tilde{\bm{f}}$, which show  deviations in the LW region due to variability in the planet's spectrum. For a GEMS target, the deviations are on the order of 0.3\% -- far above the residual scatter in the data.

We produce the binned residuals, like those shown in panel (d), for each of our different assumed planetary atmosphere scenarios. For TOI-519 b, we test both \update{full heat-redistribution} ($T_\mathrm{night}/T_\mathrm{day}=1$) and \update{moderate heat-redistribution} ($T_\mathrm{night}/T_\mathrm{day}=0.5$); we consider \update{the absence of heat-redistribution} ($T_\mathrm{night}/T_\mathrm{day}=0$) to be unrealistic for a gas-giant planet. We then invert each of our model time series data simulations from our radius-heat redistribution grid and compute the $\chi^2_\mathrm{red}$ for that model-data comparison.

\begin{figure*}
    \gridline{
    \includegraphics[width=0.5\textwidth]{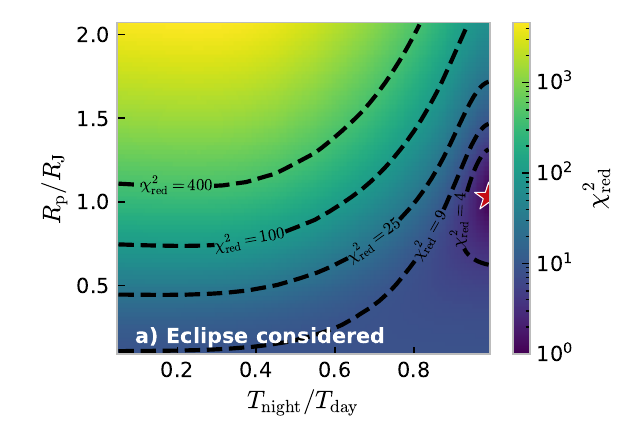}
    \includegraphics[width=0.5\textwidth]{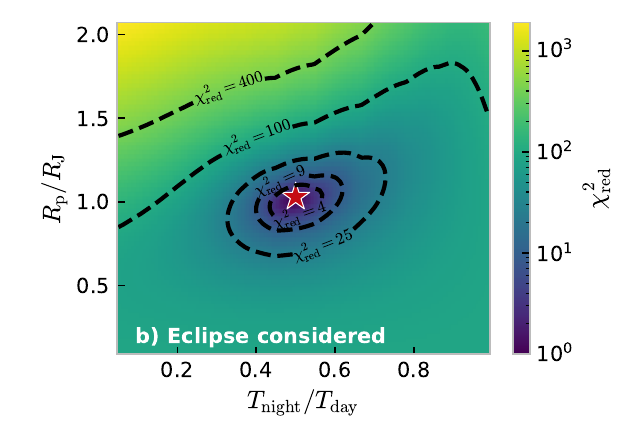}
    }
    \gridline{
    \includegraphics[width=0.5\textwidth]{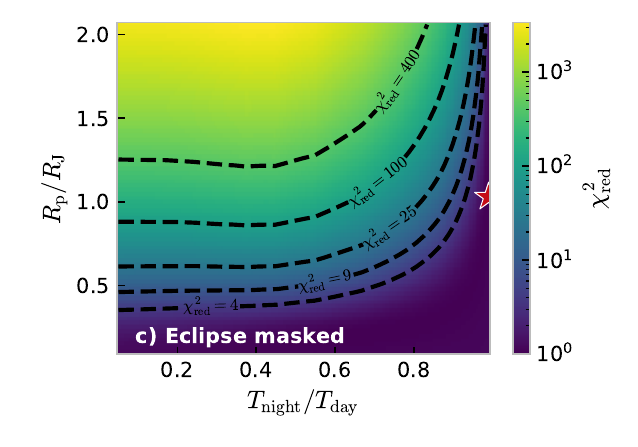}
    \includegraphics[width=0.5\textwidth]{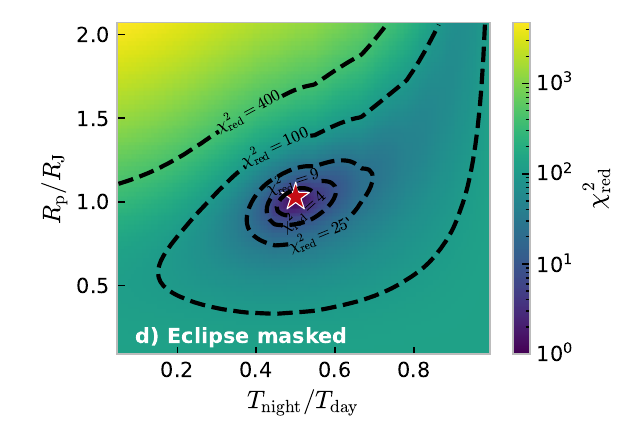}
    }
    
    \caption{
        Heat \update{re}distribution and planet\update{-}radius mismatch for TOI-519 b.
        {\bf a)} Zero variability in the planetary lightcurve, but the eclipse is considered in the analysis; by including the eclipsing epochs in the VPIE analysis the radius of the planet is tightly constrained.
        {\bf b)} Same as (a), but the $T_\mathrm{night}/T_\mathrm{day}$ of the ground-truth is 0.5.
        {\bf c)} Same as (a), but the eclipse is removed from the data before analysis. This case simulates a non-transiting planet; in the absence of an eclipse, the minimum $\chi_\mathrm{red}^2$ is degenerate between heat redistribution and planet radius. At the true $\sim R_\mathrm{J}$ radius of TOI-519 b, \update{moderate heat-redistribution} is rejected at $10\sigma$.
        {\bf d)} Same as (c), but the $T_\mathrm{night}/T_\mathrm{day}$ of the ground-truth is 0.5.
        In each panel, the star indicates the ground-truth used in calculating $\chi_\mathrm{red}^2$.
    }
    \label{fig:toi519-chirad}
    \script{toi519_inference.py}
\end{figure*}

The results are shown in Figure \ref{fig:toi519-chirad}. Each of the four panels compares a \update{simulated observation} (e.g., \update{with} $1\;R_\mathrm{Jup}$, $T_\mathrm{night}/T_\mathrm{day}=1$ in (a) and (c)) to our grid of atmosphere-radius models; regions of the parameter space with $\chi^2_\mathrm{red} > 9$ are considered to be statistically distinguishable from the data and can therefore be ruled out. Panels (a) and (b) show the results using the full data set, including the planetary eclipse, \update{whereas in panels (c) and (d), the eclipse has been masked and is therefore not included in} the VPIE analysis. 

When the planet is assumed to have full heat\update{-}redistribution, the phase curve is essentially flat and the residuals show very little signal -- a non-varying planetary flux is not \textit{exactly} correlated with the stellar variability, but it is not significantly different. Therefore both high-redistribution and \update{small radii} cannot be ruled out as the amplitude of the phase curve scales with the planet's radius and the heat recirculation efficiency. The analysis including the eclipse obviously does better in ruling out high- and low-\update{radii} because we capture the absolute flux differential during the eclipse -- but even when the eclipse is not considered, the analysis rules out scenarios with $R\gtrsim0.5\;R_\mathrm{Jup}$ and $T_\mathrm{night}/T_\mathrm{day}<0.8$. For the data that was produced using the moderate-redistribution scenario, we see that the analysis constrains the radius-heat\update{-}redistribution parameter space to a quite narrow region; in particular it finds that $0.3<T_\mathrm{night}/T_\mathrm{day}<0.7$ even when the eclipse is ignored, thereby ruling out a bare-rock (not surprising) as well as \update{Jupiter-like full heat-redistribution}. 

While the case of a warm Jupiter-sized planet around a late M-star is obviously not the most stressing case for VPIE, the results show clearly that the method can distinguish quite easily between full, moderate and zero heat-redistribution scenarios for giant planets around M-stars, even for a non-transiting target.

\subsection{Example 2: A Nearby Non-Transiting Warm Sub-Neptune in the NIR} \label{subsec:gj876}
Though transiting planets can serve as a key benchmark test for the VPIE method, the primary application of the method would be for constraining the thermal emission from non-transiting planets, especially smaller planets around the closest M-stars.  With this in mind, for our second simulated target we selected a known non-transiting planet that would be both hot enough to measure at wavelengths less than 5$\mu$m and close enough to provide a strong signal; the warm sub-Neptune GJ 876 d meets both of these criteria. 

GJ 876 d is the inner-most planet of a multi-planet system around one of the closest M-type stars to our Sun. There are currently four known planets in the system, with planet ``d'' having the lowest measured minimum mass (6.9 M$_\oplus$) and shortest period (1.9 days); the next shortest-period planet (``b'') has a period of 30 days.  The star has an effective temperature of 3294 K (M2.5V - M4V).  Our \texttt{VSPEC} model for GJ 876 d uses the full set of parameters given in Table \ref{tab:parameters}.

The GJ 876 planetary system is unique among non-transiting systems in that the known planets undergo strong gravitational interactions that have been characterized via RV. \citet{nelson2016} used dynamical models of the system to constrain the orbital parameters of each planet; we adopt their coplanar model parameters for planet ``d'', meaning that the mass and inclination of this non-transiting planet are taken to be well-constrained. \update{This constraint on inclination is convenient because it reduces the size of the model grid required to infer planetary properties, but it is not necessary to recover phase variations in general. In fact, the absolute phase curve amplitude of our GJ 876 d model at $5\;\mathrm{\mu m}$ does not change dramatically for $i>60^\circ$; it is reduced by only 12\% when the inclination is $60^\circ$ versus $90^\circ$. The effects of inclination on phase curve {\em variation} amplitude is slightly more pronounced; at $5\;\mathrm{\mu m}$, the nightside flux is 17\% of the dayside flux when $i=30^\circ$, but this drops to 1.4\% at $i=60^\circ$. This result agrees with findings by \citet[see their Section 4.8]{selsis2011}, who claim that thermal emission flux is a weak function of inclination for $i\gtrsim20^\circ$; therefore, in a significant portion of inclination-space, the planetary thermal emission signal is not expected to be significantly less than it would be for an edge-on orbit.}

The planetary $T_\mathrm{eq}$ is again based on an assumption of a geometric albedo of zero, and we use the same spot temperature and coverage as used previously; the VPIE-specific values listed in Table \ref{tab:parameters} were again determined iteratively. We again use observation parameters based on values for JWST NIRSpec/PRISM; we note that the PRISM mode may not be a realistic choice for this target due to detector saturation, but use it for easy comparison with the TOI-519 b example; we discuss possible mitigation strategies for this problem in Section \ref{subsec:inst}.

\begin{figure}
    \includegraphics[width=0.5\textwidth]{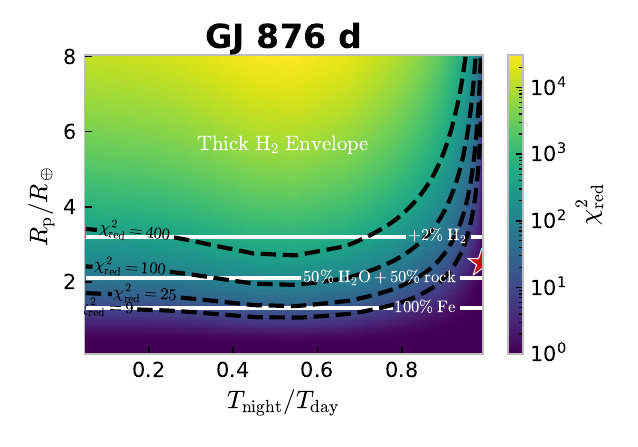}
    \includegraphics[width=0.5\textwidth]{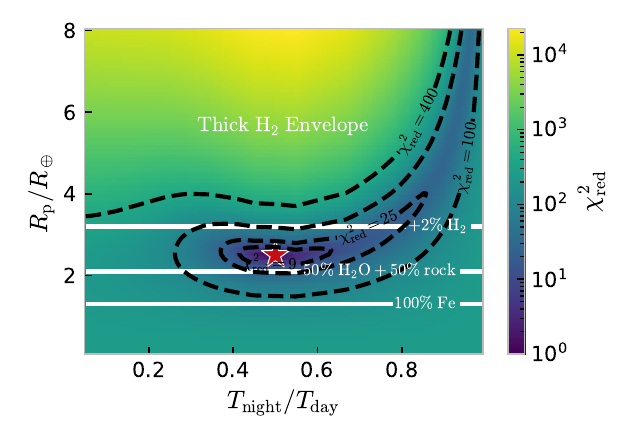}
    \includegraphics[width=0.5\textwidth]{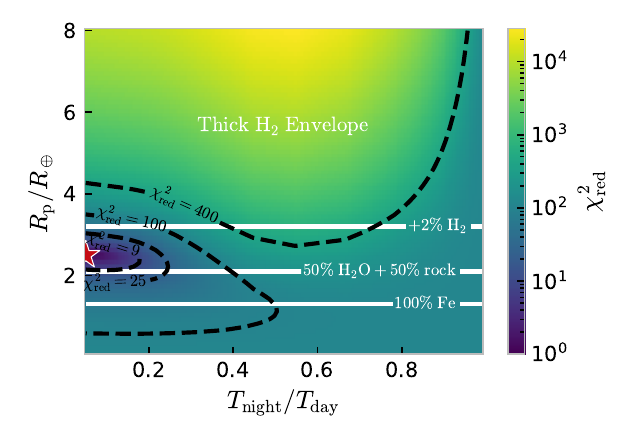}
    \caption{$\chi^2_\mathrm{red}$ model--data comparison for GJ 876 d considering three different scenarios. The star in each panel marks the ``true'' data parameters. The white curves show theoretical radius calculations based on the assumed planet mass and mass-radius curves from Figure 2 in \citet{zeng2019} for a 100\% Fe planet, a planet that is 50\% rock and 50\% water, and a planet that is equal parts rock and water with the addition of an \ce{H2} envelope that is 2\% of the total mass. As the planet's true night-day temperature ratio decreases, we are able to detect the phase curve and rule out a larger swath of radius-redistribution parameter space.}
    \label{fig:gj876-chisq}
    \script{gj876_inference.py}
\end{figure}

In Figure \ref{fig:gj876-chisq} we provide a similar $\chi^2_\mathrm{red}$ analysis as that shown in Figure \ref{fig:toi519-chirad}. In this case we test three different heat-redistribution scenarios, but since we do not have a known radius for this planet, we use a radius of $2.5\;R_\oplus$ which is typical of similar-mass planets on the sub-Neptune side of the radius gap \citep[e.g.,][]{zeng2019}. On the panels in Figure \ref{fig:toi519-chirad}, we plot the radius values that would correspond to different interior/atmosphere compositional scenarios from \citet{zeng2019}; obviously the composition is also unknown, but the delineation provides context for the types of scenarios that could be distinguished.

As expected, \update{high heat-redistribution is not well distinguished from low heat redistribution and small radius.} However, when the amplitude of the \update{observed} planetary phase curve variability becomes greater (i.e. when $T_\mathrm{night}/T_\mathrm{day}$ is small), the variable signal constrains both radius and the atmospheric regime; \update{rock- and water-dominated planets are easily distinguished from atmospheres rich in \ce{H2}.} 

These results demonstrate that for GJ 876 d, VPIE with JWST would easily be able to distinguish between \update{zero heat-redistribution} (a bare rock) and \update{moderate heat-redistribution} (a planet with an atmosphere). \update{Additionally, detectable variations in the phase-curve signal would provide constraints on the radius and therefore the interior/atmosphere composition by breaking the degeneracy between small radius and efficient heat-redistribution.}

\subsection{Example 3: A Non-Transiting Temperate Super-Earth in the MIR}\label{subsec:pcb}

In our third example, we simulate \update{observations of} the nearest exoplanet, Proxima Centauri b (PCb), using the {MIRI/LRS} instrument on JWST. With an equilibrium temperature of $243\;\mathrm{K}$ and an orbital period of 11.2 days, PCb poses a completely different challenge than the short-period warm planets that we have examined previously.  In particular, because of the low temperature, almost all of PCb's thermal flux is emitted at wavelengths longer than $5\;\mathrm{\mu m}$, with the ratio of planet-to-star emission only reaching $\sim50$ parts per million or more between 10 and $20\;\mathrm{\mu m}$ and rising at longer wavelengths.  Additionally, with such a long orbital period, the temporal binning of the data could be much coarser, which may relieve the requirement for absolute simultaneity of observations across wavelength \update{if the calibrated absolute flux is accurate enough}. 

Proxima Cen b was first discovered in 2016 using RV measurements \citep{anglada-escude2016}, and has now been shown to be part of a multi-planet system with at least 2 confirmed planets \citep{damasso2020,suarezmascareno2025}. The host, Proxima Centauri, is an M5.5V star with an effective temperature of 2900 K and a rotation period of approximately 90 days \citep{faria2022,suarezmascareno2025}. Planet b is the nearest exoplanet that is in the habitable zone, receiving about 60\% as much incident flux as Earth does. The innermost planet, d, is about twice as massive as Mars and orbits with a period about half that of planet b. In this example we only simulate planet b and the host star; we discuss the implications of and mitigation strategies for observing multi-planet systems in Section~\ref{sec:disc}. 

For this example, we use observing parameters and assumptions based on the low-resolution spectroscopy mode (LRS) on MIRI.  We chose to use the LRS mode for this analysis since it provides continuous wavelength coverage from 5 to 11~\um{} at a resolving power of $R=100$; this observing mode is similar to a longer-wavelength version of PRISM and therefore provides a reasonable comparison to our previous examples.  The other observing modes on MIRI -- the moderate-resolution spectroscopy mode (MRS) and the photometry mode -- offer wavelength coverage out to 28~\um{}, but they do not provide continuous and simultaneous spectral coverage.  This may not rule them out for VPIE-like observations, but they present challenges that are non-trivial to evaluate; we discuss this topic in more detail in Section~\ref{sec:disc}. Table \ref{tab:parameters} shows the full set of parameters used in our VSPEC model.

\begin{figure}
    \includegraphics[width=0.5\textwidth]{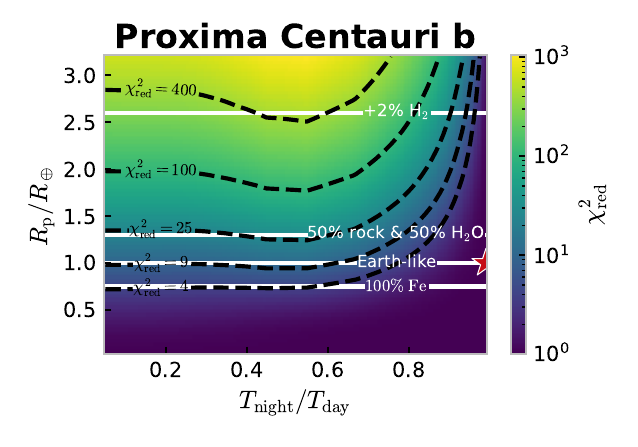}
    \includegraphics[width=0.5\textwidth]{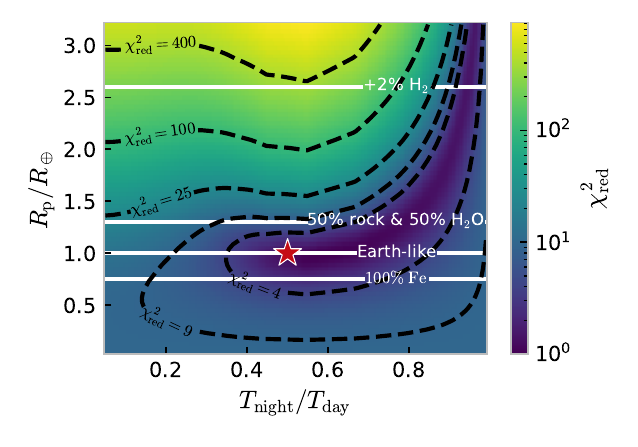}
    \includegraphics[width=0.5\textwidth]{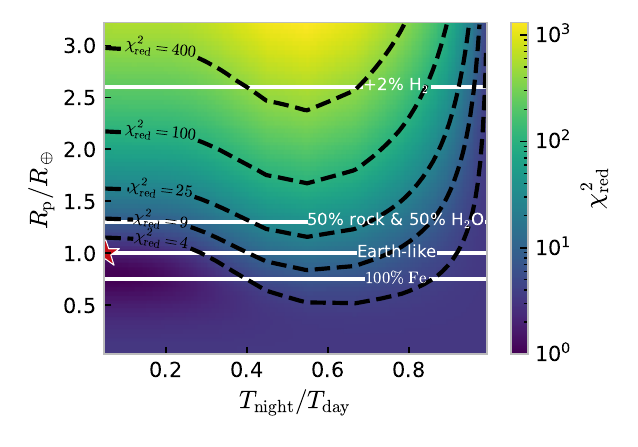}
    \caption{Same as Figure \ref{fig:gj876-chisq}, but for Proxima Centuri b observed with JWST/MIRI-LRS. The results only provide a limited constraint on the planetary radius, due primarily to the limited signal even at the long-wavelength end of the LRS spectrum.}
    \label{fig:pcb-jwst}
    \script{proxb_inference.py}
\end{figure}

Results are shown in Figure \ref{fig:pcb-jwst}. Again, we test three different heat-redistribution scenarios against a grid of models to assess the detectability of each type of atmosphere. In each panel, the simulated data is based on a $1\;R_\oplus$ planet with a night-day contrast value of $R_\mathrm{night}/R_\mathrm{day}=$ 0, 0.5, or 1. Overplotted in white are radius values for planetary compositions taken from \citet{zeng2019}. For all of the scenarios, it is difficult to distinguish atmospheres on planets with Earth-like compositions from each-other at a high statistical significance; at $R_\mathrm{p}=0.5\;R_\oplus$ virtually all heat-redistribution regimes are allowed by the data regardless of choice of true $T_\mathrm{night}/T_\mathrm{day}$.  Volatile-rich (large-radius) atmospheres are relatively well-constrained, with significant portions of the parameter space statistically excluded. The low sensitivity seen at low radii is primarily due to the signal-to-noise ratio at the longest wavelengths of MIRI LRS; At 11~\um{} even the signal of the zero \update{heat-redistribution} is only detectable for planets with larger radii. We discuss the potential use of longer-wavelength MIRI modes as well as the design requirements of an VPIE-optimized MIR instrument that would be more suitable for low-temperature targets in Section \ref{subsec:inst}.

\section{Discussion}
\label{sec:disc}
The examples provided above, while not exhaustive in any way, show that under reasonable observing conditions the VPIE method should be able to constrain the presence of an atmosphere on non-transiting planets around the nearest stars.

However, there are a number of observational challenges and methodological improvements that must be addressed before we can fully evaluate the effectiveness of VPIE for any particular observing scenario and exoplanetary target.  Below we explore these questions and provide next steps for addressing them.

\subsection{Multiple Thermal Sources in a System}
The primary challenge of using thermal emission to characterize a multi-planet system is that various combinations of radii and temperature can lead to similar levels of thermal emission flux, causing degeneracies in any inference into planetary properties. Additionally, non-planetary thermal emission features such as debris disks can produce thermal signatures that mimic planets.

The effects of multiple thermal sources in a PIE observation was first studied by \citet{mayorga2023}, who showed that breaking the multi-planet degeneracy could be accomplished by requiring the full planetary thermal emission model to be consistent with other known properties of the planet such as orbital period, mass, equilibrium temperature, and reasonable estimates for radius.  They found that thermal emission, coupled with some additional constraints on planet properties from an incomplete set of transits, could reasonably reproduce the properties of the TRAPPIST-1 planets as they have been inferred from the full set of data on that system. \citet{mayorga2023} assumed a series of integrated thermal emission measurements; the VPIE method intrinsically utilizes the same strategy by measuring a time-dependent phase curve model. Each component would also vary with a unique period, thereby facilitating the breaking of the degeneracy present in a single phase-integrated spectrum.

Dusty debris disks, on the other hand, produce thermal emission which \update{has not been observed to} vary on timescales of a planetary orbital period \citep[e.g.,][]{balog2009,hughes2018}. This essentially constant thermal signal is degenerate with thermal emission from a non-varying planet with unknown temperature and radius. However, debris disks typically emit over a broad wavelength range due to the spatially extended range of the dust distribution, so photometric measurements of a system's spectral energy distribution \update{(SED)} over a wide spectral range (i.e. a traditional infrared excess measurement) can provide important context as to the need to account for a debris disk and the potential magnitude of the signal in the VPIE wavelength range. MIR/FIR infrared excess studies of nearby M-dwarfs have found no evidence of detectable warm dust \citep{gautieriii2007}, but it should be considered as a possibility; a \update{static disk SED} can be added to planetary phase curves during parameter retrievals if needed.


\subsection{Impact of Additional Complexity in Planetary and Stellar Properties}
Each of the examples provided above simulated extremely simplified star/planet systems, and future work should consider more complex models for both stellar and planetary components. The planet models used here were simple day-night heat redistribution functions that depend only on the equilibrium temperature, day/night temperature ratio, and planetary radius. For example, realistic planetary models include contributions from atmospheric absorbers and aerosols, the potential for a non-zero albedo, more realistic heat transport \update{including tides and other source of internal heating}, and the impact of an uncertain orbital inclination and planetary obliquity. A realistic atmosphere may be optically thick due to clouds, hazes, or even high concentrations of gas, leading to cooler emission for higher up that diminishes the efficacy of thermal emission measurements. Spatial distributions of clouds complicate the simple $T_\mathrm{night}/T_\mathrm{day}$ parameterization employed above; \citet{hammond2025}, for example, found that some nearby rocky planets might emit more on their night sides due to windows through an otherwise\update{-}thick cloud layer.

Additionally, we have made fairly simple assumptions about the host star. In each example, we assumed a star with a single photosphere temperature and a single spot temperature. While the visible fraction of the surface covered by spots was allowed to change as the star rotated, spots on a real star may change temperature as they evolve and are not homogeneous \citep[e.g. sunspots, see][]{solanki2003}. Other sources of stellar inhomogeneity include hot faculae, and stellar flares and other stochastic events may occur over a range of amplitudes and frequencies \citep[e.g.][]{spruit1976,topka1997,magic2014,gordon2020,hovis-afflerbach2022,gao2022}.

The solution to the planetary complexity problem is both \update{straight-forward} and expensive: to infer more realistic planets we need to compute the residual \update{deviation} against more realistic (and expensive) planetary models. To infer more planetary properties (i.e. albedo or inclination) we need to generate higher-dimensional model grids. Both of these endeavors are feasible, but are not necessary for the proof-of-concept simulations we have provided.

Stellar complexity, on the other hand, directly drives the number of basis vectors needed to adequately describe an observation. Additional basis vectors -- so long as the number is much less than the number of epochs observed -- are not a problem. The addition of each basis removes a small amount of information about the planet. Stellar complexity becomes a problem when the required number of basis vectors becomes very high. For example, the intrinsic spectra of stellar flares change significantly over a flare's lifetime \citep{howard2025}; each observed epoch of a flaring star may contain a flare contribution with a unique spectrum, in which case the number of basis vectors would equal the number of epochs. One potential strategy to mitigate the effects of flares is to simply \update{mask} epochs with known flare activity, but this may not be feasible if flares are ubiquitous; on the other hand, the SED of flares may be easily modeled at long wavelengths. Further work (and improved knowledge of flare SEDs) is necessary to better understand these factors.
\update{Flares have also been found to alter the surfaces of their host star, providing additional sources of potential complication. However, \citet{vasilyev2025} suggest that post-flare changes to the stellar spectrum are primarily due to the sudden disappearance of spots, which would mean that no additional basis vectors are required if flares are adequately masked.}

\subsection{Instrument Suitability}\label{subsec:inst}
In the above examples we have assumed somewhat idealized analogs to JWST instruments. In this section we discuss limitations that would make these observations challenging in practice. A requirement of VPIE (and PIE in general) is that SW and LW spectra are observed simultaneously. Most instrument modes that observe a wide-enough simultaneous wavelength range utilize a lower spectral resolution (to conserve detector real estate) and therefore also saturate more quickly. This is not an issue for TOI-519, which is sufficiently small and distant to avoid saturating the JWST/NIRSpec PRISM mode.  However, GJ 876 is much too close at $<4.7\;\mathrm{pc}$ to be observed with standard read-out schemes for PRISM, and a similar problem exists for Proxima Cen using MIRI/LRS. But this challenge of observing bright exoplanet targets with various JWST observing modes is well-known, and new innovative observing strategies continue to be explored by the mission and instrument science teams.  For example, a new mode for utilizing both the SW and LW channels of the NIRCam instrument simultaneously for bright-object spectroscopy by utilizing the dispersed Hartmann sensor (DHS) on the SW channel is now being deployed \citep{schlawin2017}. 

Even when brightness limits are ignored, limited spectral coverage at thermal wavelengths relevant to cooler, potentially-habitable planets can make characterization of these targets difficult. In Section \ref{subsec:pcb}, we showed that the atmosphere of a rocky PCb would be poorly constrained with JWST/MIRI LRS; planet/star contrast would much more favorable with an instrument that captures longer wavelengths. Unfortunately, there are no modes on MIRI that fully cover the MIR wavelength range (5 -- 28~$\mu$m) simultaneously. The LRS mode fully covers 5 -- 11~$\mu$m but does not extend farther, while the MRS (moderate-resolution spectroscopy) mode covers the full 5 -- 28~$\mu$m but does so with four non-contiguous wavelength channels; additionally, the integral field units (IFUs) utilized by the MRS mode make it challenging for high-precision time series observations. Asynchronous photometric measurements using multiple filters could also be attempted, but this would also face saturation issues for bright targets and would result in an inhomogeneous data set that could pose additional problems in analysis.

Of course, these challenges could be mitigated through the deployment of a NEW instrument which is specifically designed for PIE time-series observations. Unfortunately upgrading the instrumentation on JWST does not appear to be a feasible option, but several concepts for future missions could incorporate such an instrument. \update{The dedicated Explorer-class mission concept called MIRECLE (M-star Infrared Exoplanet Climate Explorer) \citep{staguhn2019,mandell2022} would be optimized to utilize VPIE for a survey of non-transiting rocky planets around nearby M-stars}, and instrumentation for future Probe- and flagship-class missions such as PRIMA and the Far-IR Surveyor \update{could also be designed to accomodate VPIE observations}. A low-resolution, high-precision spectrometer with saturation limits designed for the nearest M-stars would open dozens of nearby rocky planets for atmospheric characterization.

\begin{figure*}
    \gridline{
        \includegraphics[width=0.3\textwidth]{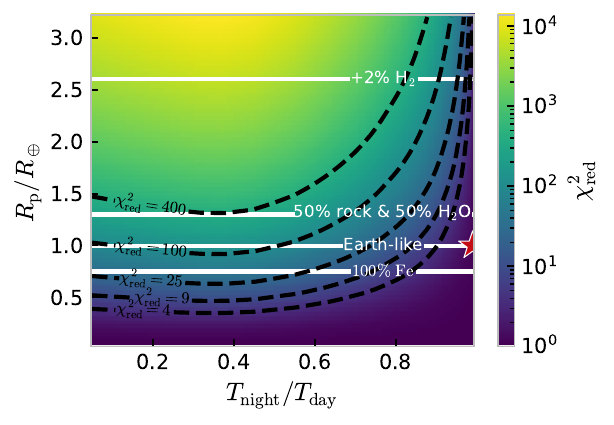}
        \includegraphics[width=0.3\textwidth]{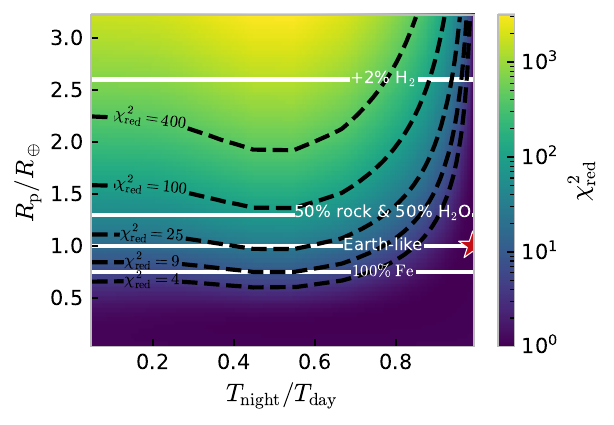}
        \includegraphics[width=0.3\textwidth]{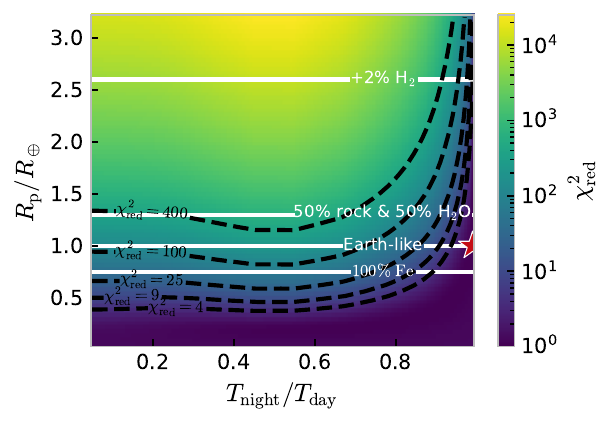}
    }
    \gridline{
        \includegraphics[width=0.3\textwidth]{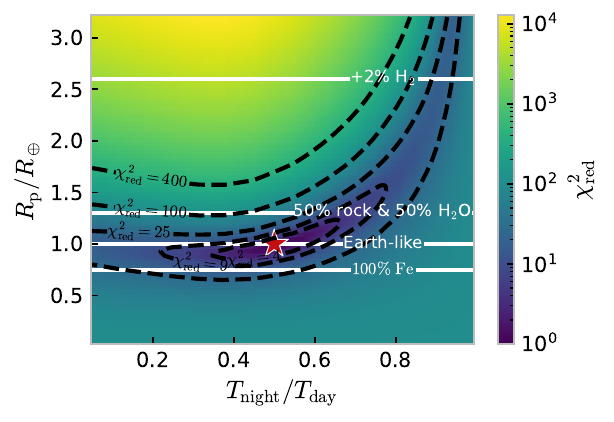}
        \includegraphics[width=0.3\textwidth]{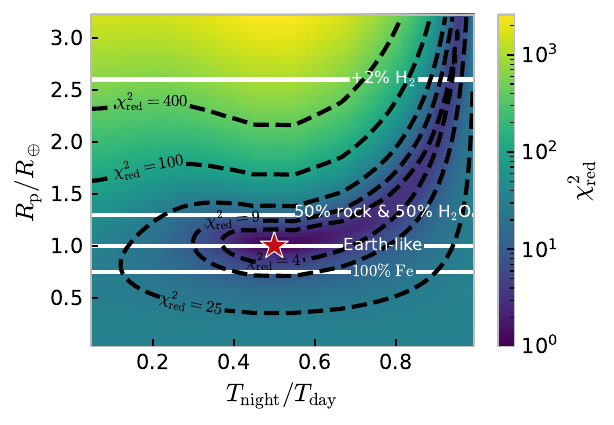}
        \includegraphics[width=0.3\textwidth]{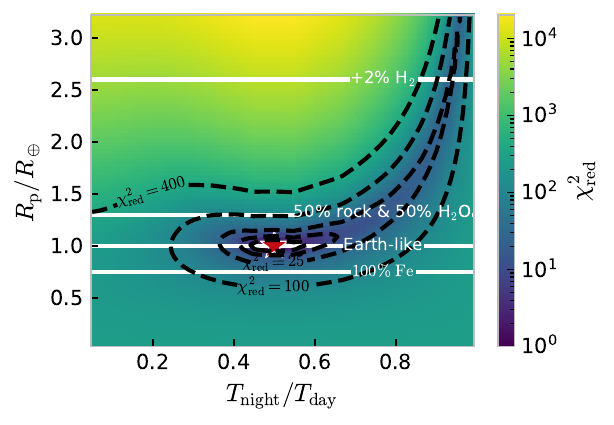}
    }
    \gridline{
        \includegraphics[width=0.3\textwidth]{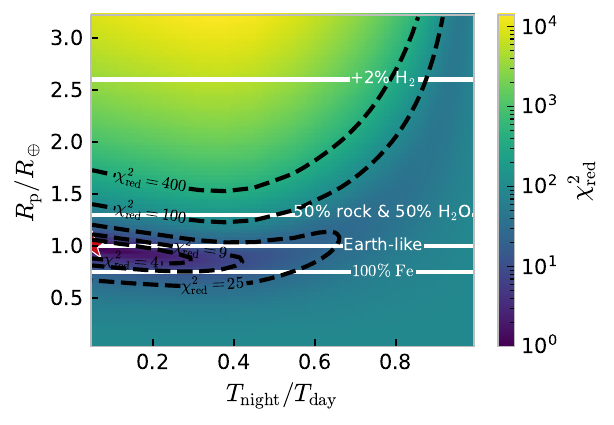}
        \includegraphics[width=0.3\textwidth]{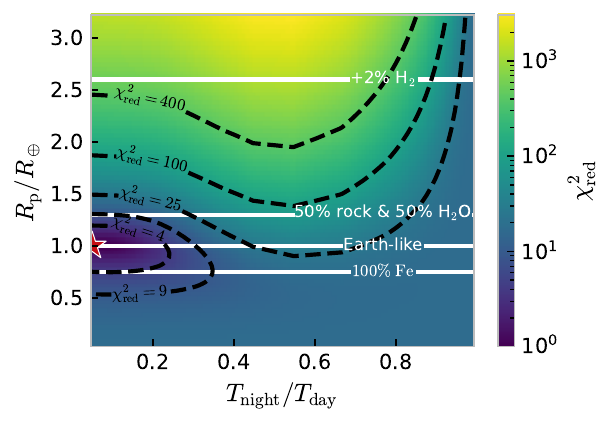}
        \includegraphics[width=0.3\textwidth]{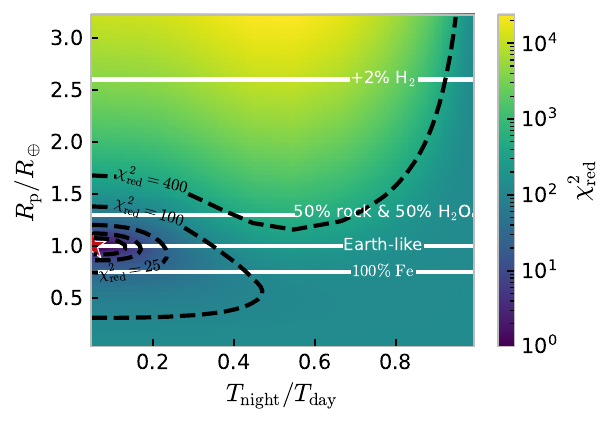}
    }
    \caption{Proxima Centauri b atmosphere detectability for a MIRECLE-like mission with three different apertures. The left panels simulate an observation with a 0.5 m telescope taken over 16 orbits of PCb ($\sim$ 6 months). This observation length was chosen so that the photon noise would be comparable to that of a 2 m telescope observing for a single orbit, shown in the center column. The 16-orbit observation was phase-folded prior to analysis. The right column panels show a simulation of a MIRECLE-like mission with the same aperture as JWST\update{, emulating a future FIR Surveyor}. Enhanced detectability of phase-variations in this atmosphere compared to Figure \ref{fig:pcb-jwst} are due to the expanded wavelength range of the MIRECLE instrument concept. 
    }
    \label{fig:mirecle}
    \script{mirecle_inference.py}
\end{figure*}

To investigate the effects of optimized instrument properties, we have simulated observations of PCb with three versions of a future observatory in Figure \ref{fig:mirecle}. The first emulates a low-cost Small Explorer-class mission with a 0.5 m aperture mirror. After observing 16 orbits of PCb over approximately six months \update{the phase curve is phase-folded at Proxima Centauri b's orbital period; Phase-folding the raw data is necessary because it}
exhibits a high amount of scatter, which introduces correlated noise to the residual and masks the planet's signal. Essentially, the error term ($\bm{\epsilon}$) in equation \ref{eq:def-residuals} \update{would otherwise} dominate over the signal term ($\bm{\delta}$), \update{and in fact, we found that our analysis performed poorly when applied to the raw, unfolded data}.

\update{Combining a very long observation and the phase-folding technique, this small-mirror telescope performs very well. In every case, we can place limits on radius and atmosphere which are even more strict than those found using a 2-meter aperture and a single orbit, which is an apt comparison considering each observation would receive the same number of photons. In practice, instrument stability could pose a major challenge over the course of a 6 month observation, but these results are nonetheless encouraging that PCb could be characterized with a low-cost, dedicated observatory.}

For the 2-meter aperture telescope similar to the mission architecture described in \citet{staguhn2019} and \citet{mandell2022}, despite only observing for one orbit the data produces \update{meaningful} constraints on both day-night temperature ratio and radius. The larger collecting area is capable of constraining the atmospheric heat redistribution at a $>3\;\sigma$ level for rocky, Earth-like planets and even small metal-dominated planets.

Our last set of simulations emulate a future flagship infrared mission, such as the Far-IR Surveyor mission that was recommended by the Astro2020 Decadal Survey \citep{committeeforadecadalsurveyonastronomyandastrophysics2020astro20202023}. We use the collecting area of JWST along with the instrument parameters described for MIRECLE. By comparing to the simulation of a JWST/MIRI-LRS observation of PCb in Figure \ref{fig:pcb-jwst}, we see how critical observing over a wide wavelength range is to VPIE; MIRI-LRS's 5 -- 11~\um{} coverage allows the instrument to place up to $4\;\sigma$ constraints on Earth-like models -- but with MIRECLE's 1 -- 18~\um{} it is $>10\;\sigma$.

Though all \update{the issues discussed in this section} pose challenges to the practical applicability of the method using current observing assets, none of them are inherently insurmountable with either more sophisticated analysis techniques or more suitable instrumentation in the future. We plan to examine the implications and best-practices to address these issues in future work.

\section{Conclusions}\label{sec:conclusion}
In this paper we present a new method for characterizing non-transiting planets around the nearest low-mass stars, which we call the Variable Planetary Infrared Excess (VPIE) technique. By modeling stellar variability using only short-wavelength spectra and treating the time series as a vector space, VPIE isolates long-wavelength residuals that are encoded with phase-dependent planetary emission. This approach is intrinsically agnostic to stellar atmosphere models and is therefore less susceptible to systematic errors arising from imperfect stellar spectral predictions.

Using simulated JWST observations of representative exoplanetary systems, we have demonstrated that VPIE can recover meaningful constraints on planetary radius and atmospheric heat redistribution under realistic observing conditions. For the warm giant planet TOI-519 b, the method robustly distinguishes between different circulation regimes, even in the absence of secondary eclipses. For the nearby sub-Neptune GJ 876 d and the closest terrestrial exoplanet Proxima Centauri b, VPIE can rule out volatile-rich atmospheres and strong day–night temperature contrasts, although its performance \update{for observations of cooler planets with JWST is increasingly limited by sub-optimal wavelength coverage}.

These results emphasize the utility of VPIE for characterizing the large population of nearby, non-transiting planets that are inaccessible to traditional eclipse- and transit-based techniques. As radial velocity and astrometric surveys continue to identify planets around the nearest low-mass stars, methods capable of inferring atmospheric properties for non-transiting planets will play an increasingly important role in exoplanet science.

We have also identified several areas requiring further investigation. More realistic treatments of stellar heterogeneity, flaring, and multi-planet systems are needed to assess the robustness of the method in complex environments. In addition, instrumental constraints—particularly those related to wavelength coverage, simultaneity, and saturation—will strongly influence the applicability of VPIE to specific targets; we lay out the performance potential of VPIE observations with a new custom-built instrument for future MIR/FIR observatories.

Future work will focus on validating VPIE with archival and forthcoming JWST time-series observations, further exploring atmospheric inference strategies to extract more detailed planet properties, and guiding instrument design for next-generation infrared facilities. In particular, dedicated instruments optimized for broad, simultaneous wavelength coverage and high photometric stability would substantially enhance the scientific return of VPIE-style analyses.

In summary, the VPIE technique provides a flexible and physically motivated framework for isolating planetary thermal emission in combined-light spectroscopic time series. By leveraging correlated stellar variability at short wavelengths, it enables phase-resolved characterization of both transiting and non-transiting exoplanets. This capability expands the accessible parameter space for atmospheric studies and represents a significant step toward a more comprehensive census of planetary climates in the solar neighborhood.

\begin{acknowledgments}
The authors thank R. G. Martin for valuable editorial assistance and thoughtful mentorship, which greatly contributed to the clarity of the mathematics in this manuscript and to the professional development of T. M. Johnson. They also thank K. Stevenson for helpful discussions about PIE and VPIE concepts, and acknowledge support from the Sellers Exoplanet Environments Collaboration, which is funded through the Planetary Science Division at NASA. T.M.J. also acknowledges support from NASA through grants 80NSSC21K0395 and 80NSSC19K0443. \update{The authors thank R. K. Kopparapu for conversations about the population of nearby exoplanets which motivated Figure \ref{fig:nearby_mdwarfs}.}

\end{acknowledgments}
\section*{Open Science}
\update{This manuscript was prepared using the open-science software \href{https://show-your.work/en/latest/intro/}{\showyourwork} \citep{luger2021}, making the article completely reproducible. The \GitHubIcon icon next to each figure links to the source code which was used to generate it. The source code to compile this document and create the figures from scratch is available on GitHub\footnote{\ghurl}. Additionally, a copy of the repository has been deposited on Zenodo\footnote{\zenodourl}}.

\software{
\update{ASDF \citep{greenfield2015,greenfield2022,brettgraham2025}}\\
\update{Astropy \citep{astropycollaboration2013,astropycollaboration2018,astropycollaboration2022}}\\
\update{NumPy \citep{harris2020}}\\
\update{Matplotlib \citep{hunter2007}}\\
\update{pandas \citep{mckinney2010,thepandasdevelopmentteam2025}}\\
\update{PSG/GlobES \citep{villanueva2018,kofman2024}}\\
\update{SciPy \citep{virtanen2020}}\\
\update{\texttt{vpie-rs} \citep{vpie-rs-0.2.0}}
\update{\texttt{VSPEC}, GridPolator, \& \texttt{libpypsg} \citep{johnson2025}}
}

\appendix

\section{Mathematical description of VPIE}
\label{ap:math-desc}

In this section we mathematically define many of the processes that Section \ref{sec:vpie} describes in plain language.

\subsection{Matrix description of an observation}
\label{ap:mat-desc-obs}

Here we define a reduced VPIE dataset in matrix form. After a spectral phase curve has been put through standard data-reduction procedures, we are left with two matrices: $\mathbf{F}$ -- the flux from the system, and $\mathbf{\Sigma}$ -- uncertainties which we assume are dominated by photon noise. Both are $m\times n$ matrices, where $m$ is the number of epochs and $n$ is the number of wavelength channels. If $n'$ is the number of SW wavelength channels, then we can define the SW data to be the $m\times n'$ matrix $\mathbf{F}_\text{SW}$. We will also use the subscripts $*$ and $\text{p}$ to represent flux from the star and planet, respectively. By definition of the two wavelength regimes, we can say that
\begin{equation}
    \label{eq:fnir}
    \mathbf{F}_\text{SW} = \mathbf{F}_{\text{SW},*}\, ,
\end{equation}
and
\begin{equation}
    \mathbf{F}_\text{LW} = \mathbf{F}_{\text{LW},*} + \mathbf{F}_{\text{LW},\text{p}}\, .
\end{equation}

We also define a model of the observation, $\tilde{\mathbf{F}}$, which is found by minimizing the squares of an error matrix defined
\begin{equation}
    \label{eq:def-E}
    \mathbf{E} \equiv \mathbf{F}_* - \tilde{\mathbf{F}}_*\,.
\end{equation}

By equation (\ref{eq:fnir}), at short wavelengths this means
\begin{equation}
    \mathbf{F}_\text{SW} - \tilde{\mathbf{F}}_\text{SW} = \mathbf{E}_\text{SW}\, .
    \label{eq:nir-error}
\end{equation}

\subsection{The meaning of LW residuals}\label{ap:residuals}
The long-wavelength counterpart of equation (\ref{eq:nir-error}) is
\begin{equation}
    \mathbf{F}_\text{LW} - \tilde{\mathbf{F}}_\text{LW} = \mathbf{E}_\text{LW} + \mathbf{\Delta}\, ,
    \label{eq:mir-error}
\end{equation}
where $\mathbf{\Delta}$ is the part of the residual attributed not to noise, but to planetary flux. It is important to note that the $\mathbf{E}_\text{SW}$ minimization procedure described in Section \ref{ap:compute-A} also minimizes $\mathbf{E}_\text{LW}$, but has no effect on $\mathbf{\Delta}$. Section \ref{ap:compute-model} describes a method to compute $\tilde{\mathbf{F}}$ from $\mathbf{F}$, and so the residual $\mathbf{E} + \mathbf{\Delta}$ is a measurable quantity.

We later discuss $\tilde{\mathbf{F}}$ as the product of a coefficient matrix $\mathbf{A}$ and a basis matrix $\mathbf{B}$ (see eq. \ref{eq:f-eq-ab}). Since $\mathbf{B}$ is composed of spectra, we can write it as a sum of stellar and planetary components:
\begin{equation}
    \mathbf{B} = \mathbf{B}_* + \mathbf{B}_\text{p}\, ,
\end{equation}
and so $\tilde{\mathbf{F}}$ can be written
\begin{align}
    \tilde{\mathbf{F}} & = \mathbf{A}\, (\mathbf{B}_* + \mathbf{B}_\text{p})\, , \nonumber \\
    & = \tilde{\mathbf{F}}_* + \tilde{\mathbf{F}}_\text{p}\, .
\end{align}

Rewriting equations (\ref{eq:nir-error} \& \ref{eq:mir-error}) as a single statement, we see
\begin{equation}
    \mathbf{E} + \mathbf{\Delta} = \mathbf{F}_* - \tilde{\mathbf{F}}_* + \mathbf{F}_\text{p} - \tilde{\mathbf{F}}_\text{p}\, .
\end{equation}

Given the definition of $\mathbf{E}$ in equation (\ref{eq:def-E}), we see that
\begin{align}
    \mathbf{\Delta} &= \mathbf{F}_\text{p} - \tilde{\mathbf{F}}_\text{p}\, , \\
    & = \mathbf{F}_\text{p} - \mathbf{A}\mathbf{B}_\text{p}\, . \label{eq:calc-glob-delta}
\end{align}

Specifically, $\mathbf{\Delta}$ is an observable that can be predicted by a planetary model. The following sections aim to minimize $\mathbf{E}$ so that the residual $\mathbf{F} - \tilde{\mathbf{F}} \approx \mathbf{\Delta}$, and we can use it to make inferences about properties of the planet.

\subsection{Computing a model from a basis}
\label{ap:compute-model}
In this section we aim to compute a reconstructed observation model, $\tilde{\mathbf{F}}$ given a basis. Recall that the original observation, $\mathbf{F}$, contains $m$ rows representing epochs and $n$ columns representing wavelength channels.

Suppose we are given a basis (chosen in a way described in Appendix \ref{ap:choose-basis}) of $q$ spectra. We can represent this basis with a set, $s$, of size $q$ that contains the row indices of each spectrum in $\mathbf{F}$. For convenience, the elements of $s$ are ordered, but the ordering method is unimportant. We can define a basis matrix, $\mathbf{B}$, such that the rows of $\mathbf{B}$ are the basis spectra. In other words, $\mathbf{B}$ is a $q\times n$ matrix with elements
\begin{equation}
    B_{kj} = F_{s_kj} \quad \text{for}\quad k=1,\,2,\,\dots,\,q\,.
\end{equation}

We can also define a coefficient matrix, $\mathbf{A}$, which has shape $m\times q$. This is the matrix that is found via least-squares to best approximate the original observation. The model observation can be written
\begin{equation}
    \tilde{\mathbf{F}} = \mathbf{A}\mathbf{B}\, .
    \label{eq:f-eq-ab}
\end{equation}

\subsubsection{Computing the coefficient matrix}
\label{ap:compute-A}
Consider the following row vectors: $\bm{f}$ (a row of $\mathbf{F}$), $\bm{\sigma}$ (a row of $\mathbf{\Sigma}$), $\bm{\epsilon}$ (a row of $\mathbf{E}$), and $\bm{a}$ (a row of $\mathbf{A}$). If these are all the $i$th row of their respective matrix, then they correspond to the same epoch. The procedure here is simple: given $\mathbf{B}$, $\bm{f}$, and $\bm{\sigma}$, perform weighted least squares to compute $\bm{a}$ which minimizes the squares of $\bm{\epsilon}$. We have dropped the SW subscript in this subsection, but it is important that only SW spectra are used.

First, we define a diagonal weight matrix, $\mathbf{W}$, where the diagonal elements are the inverse of the variance, i.e. $W_{jk} = \delta_{jk}/(\sigma_j\sigma_k)$, where $\delta_{jk}$ is the Kronecker delta. Now, $\bm{a}$ is the vector that produces the least-squares solution to the equation
\begin{equation}
    \mathbf{W}\bm{f}^\top = \mathbf{W}\mathbf{B}\bm{a}^\top\,.
\end{equation}

Note that if the covariance between wavelength channels is known, then $\mathbf{W}$ can be generalized to be the non-diagonal precision matrix. The generalized least-squares solution is the same:
\begin{equation}
    \label{eq:a_transpose}
    \bm{a}^\top = (
        \mathbf{B}\mathbf{W}^2\mathbf{B}^\top
    )^{-1} \mathbf{B}\mathbf{W}^2 \bm{f}^\top\, .
\end{equation}

In equation (\ref{eq:a_transpose}) we have computed a row of the coefficient matrix $\mathbf{A}$. The full matrix is then
\begin{align}
    \mathbf{A} &= \begin{bmatrix}
        \bm{a}_1 \\ \bm{a}_2 \\ \vdots \\ \bm{a}_i \\ \vdots \\ \bm{a}_m 
    \end{bmatrix}\, , \\
    \bm{a}_i &= (
        (
        \mathbf{B}\mathbf{W}_i^2\mathbf{B}^\top
    )^{-1} \mathbf{B}\mathbf{W}_i^2 \bm{f}_i^\top
    )^\top\, ,
\end{align}
where the index $i$ indicates the respective epoch or time index. Additionally, note that computing $\mathbf{A}$ requires inverting the $q\times q$ matrix $\mathbf{B}\mathbf{W}_i^2\mathbf{B}^\top$ $m$ times. It may be more computationally efficient to use a time-averaged weight matrix $\langle\mathbf{W}\rangle$ to compute this quantity once and cache its value for future use.

\subsection{Choosing a spectral basis}
\label{ap:choose-basis}
The goal of this section is to find the smallest set of linearly-independent spectra that together compose a vector space containing all the spectra in our observation. This basis is defined by the optimized set of basis indices, $\hat{s}$ (opposed to a general non-optimized index set $s$), which can be used to compute the basis matrix $\mathbf{B}$. $\hat{s}$ is the set $s$ which minimizes some model selection function (e.g. BIC or AIC). We will use BIC here because of its simplicity. BIC requires we compute a maximized likelihood function, $\hat{L}$, which is the probability of observing $\mathbf{F}$ given the model $\tilde{\mathbf{f}}$ and uncertainties $\mathbf{\Sigma}$. The BIC is
\begin{equation}
    \text{BIC} = q\ln{(mn)} - 2\ln{(\hat{L})}\, ,
\end{equation}
and the maximized likelihood can be computed
\begin{equation}
    \hat{L} = \prod_{i=1}^m \prod_{j=1}^n \frac{1}{\sqrt{2\pi \Sigma_{ij}^2}} \exp{\left (-\frac{(F_{ij} - \tilde{F}_{ij})^2}{2\Sigma_{ij}^2}\right )}\, .
\end{equation}

The BIC is then written
\begin{equation}
    \label{eq:bic2}
    \text{BIC} = q\ln{(mn)} + \sum_{ij} \left (\ln{(2\pi\Sigma_{ij}^2)} + \frac{(F_{ij} - \tilde{F}_{ij})^2}{\Sigma_{ij}^2}\right )\, .
\end{equation}

Note that, for a single dataset, $m$, $n$, $\mathbf{F}$, and $\mathbf{\Sigma}$ are constant, and $q$ and $\tilde{\mathbf{F}}$ determined only by our choice of $s$. Therefore, there exists a function $\text{BIC}(s)$ which can be used to compare any choice of basis $s$ to any other choice $s'$.

\subsection{Basis search algorithms}

All that is left now is to test various values of $s$, and choose the one for which BIC is minimized. There are many ways to do this. The simplest is to start at low values of $q$ and step through each possible value of $s$ sequentially, finding a local minimum for each value of $q$. When increasing $q$ no longer leads to decreased BIC, the search is stopped and the previous minimum is returned as the solution.
\begin{center}
\begin{lstlisting}[
    label={ls:soln1}, caption={Minimizing BIC though an exhaustive search at low $q$}
]
minima = {}
for q in 0..m-1:
    sets = combos(0..m-1, q)
    vals = []
    for s in sets:
        vals.append(bic(s))
    imin = argmin(vals)
    if q > 1:
        if vals[imin] < minima[q-1]["val"]:
            minima[q] = {
                "val":vals[imin],
                "s":sets[imin]
            }
        else:
            return minima[q-1]["s"]
    else:
            minima[q] = {
                "val":vals[imin],
                "s":sets[imin]
            }
\end{lstlisting}
\end{center}

This method, shown as pseudocode in Listing \ref{ls:soln1} is exhaustive, but quickly becomes computationally expensive. There are $m\choose q$ combinations for each loop to consider, so the time complexity of each loop iteration is $\mathcal{O}(m^q)$.

A less expensive method is to find local minimum with $q=1$, and then add a single index at a time as $q$ is increased. This will not necessarily find the global minimum of the BIC, but searches higher values of $q$ much faster as the time complexity is $\mathcal{O}(m)$. A pseudocode implementation is shown in Listing \ref{ls:sequential}.

\begin{center}
\begin{lstlisting}[
    label={ls:sequential},caption={Minimizing BIC by searching for the next best index for each value of $q$}
]
s = []
vals = []
for _ in 1..m-1:
    _is = []
    _vals = []
    for i in 1..m-1 if i not in s:
        _s = s + [i]
        _is.append(i)
        _vals.append(bic(_s))
    imin = argmin(_vals)
    s.append(_is[imin])
    vals.append(_vals[imin])
q_best = argmin(vals) + 1
s = s[:q_best-1]
return s
\end{lstlisting}
\end{center}

\bibliographystyle{aasjournal}
\bibliography{syw,pie,output/nea}
\end{document}